\documentclass[preprint,floatfix,epsfig]{revtex4}

\usepackage{amssymb, amsmath,amsthm}    

\usepackage{custom_defn}

\usepackage{subfigure}
\usepackage{graphicx} 
\usepackage{dcolumn} 
\usepackage{amssymb}
\usepackage{amsmath}
\usepackage{bm}
\usepackage[dvips]{color}
\usepackage{times}
\usepackage{graphicx}
\usepackage{ulem}
\usepackage{caption}
\usepackage{tabularx}
\usepackage{feynmp} \unitlength=1mm
\usepackage{slashed}
\usepackage{slashbox}
\usepackage{appendix}

\usepackage{amstext}
\usepackage{array}

\usepackage{bbm}

\allowdisplaybreaks 

\renewcommand\theequation{{\color{blue}\arabic{equation}}}
\newcommand{\dd}{\mathrm{d}}

\newcommand{\s}{{\color{blue}s}}

\newcommand{\bqa}{\begin{eqnarray}} 
\newcommand{\eqa}{\end{eqnarray}}
\newcommand{\nn}{\nonumber \\}

\definecolor{new_color}{RGB}{50,155,0}

\newcommand{\f}{\frac}

\newcommand{\p}{\partial}

\newcommand{\eqaln}[1]{\begin{align}#1\end{align}}

\def\XXint#1#2#3{{\setbox0=\hbox{$#1{#2#3}{\int}$}
\vcenter{\hbox{$#2#3$}}\kern-.5\wd0}}

\def\veps{{\varepsilon}}

\def\[{\left[}
\def\]{\right]}
\def\({\left(}
\def\){\right)}

\def\s{\sigma}

\def\g{\gamma}

\def\({\left(}
\def\){\right)}
\def\[{\left[}
\def\]{\right]}

\def\<{\langle}
\def\>{\rangle}

\def\i2{\frac{i}{2}}

\def\spi{\relax{\rm \pi\kern-0.5em /}}
\def\sA{\relax{\rm A\kern-0.5em /}}
\def\sp{\relax{\rm p\kern-0.5em /}}
\def\sd{\relax{\rm \d\kern-0.5em /}}
\def\sk{\relax{\rm k\kern-0.5em /}}
\def\sn{\relax{\rm n\kern-0.5em /}}
\def\sl{\relax{\rm l\kern-0.5em /}}
\def\sP{\relax{\rm P\kern-0.7em /}}
\def\sBethe{\relax{\rm \Bethe\kern-0.5em /}}

\DeclareMathOperator{\arccot}{arccot}

\begin{document}

\title{Exact critical exponents for the antiferromagnetic quantum critical metal
 in two dimensions  }

\author{
Andres Schlief, Peter Lunts and Sung-Sik Lee\\ 
\vspace{0.3cm}
{\normalsize{Department of Physics $\&$ Astronomy, 
McMaster University, Hamilton, ON L8S 4M1,Canada}}\\
{\normalsize{Perimeter Institute for Theoretical 
Physics, Waterloo, ON N2L 2Y5, Canada}}\\
}

\date{\today}

\begin{abstract}

{\bf
Unconventional metallic states 
which do not support well defined single-particle excitations
can arise near quantum phase transitions 
as strong quantum fluctuations of incipient order parameters
prevent electrons from forming coherent quasiparticles.
Although antiferromagnetic phase transitions occur commonly in correlated metals,
understanding the nature of the strange metal realized at the critical point in layered systems
has been hampered by a lack of reliable theoretical methods that take into account
strong quantum fluctuations.
We present a non-perturbative solution to the low-energy theory
for the antiferromagnetic quantum critical metal in two spatial dimensions.
Being a strongly coupled theory, 
it can still be solved reliably in the low-energy limit
as quantum fluctuations are organized by 
a new control parameter that emerges dynamically. 
We predict the exact critical exponents 
that govern the universal scaling of physical observables
at low temperatures. 
}

\end{abstract}

\maketitle


\newpage

\section{Introduction} 

One of the cornerstones of condensed matter physics is Landau Fermi liquid theory, 
according to which quantum many-body states of interacting electrons 
are described by largely independent quasiparticles in metals\cite{LFL}.
In Fermi liquids, 
the spectral weight of an electron is sharply peaked at a well defined energy 
due to the quasiparticles with long lifetimes.
On the other hand, exotic metallic states beyond the quasiparticle paradigm 
can arise near quantum critical points,
where quantum fluctuations of collective modes driven by the uncertainty principle
preempt the existence of well defined single-particle excitations\cite{PhysRevB.14.1165,PhysRevB.48.7183,RevModPhys.79.1015,RevModPhys.73.797}.
In the absence of quasiparticles, 
many-body states become qualitatively different from 
a direct product of single particle wavefunctions.
Due to strong fluctuations near the Fermi surface,
the delta function peak of the electron spectral function 
is smeared out,  leaving a weaker singularity behind.
The resulting non-Fermi liquids 
exhibit unconventional power-law dependences
of physical observables on temperature and probe energy\cite{PhysRevB.78.035103}.  
A primary theoretical goal is 
to understand the universal scaling behavior of the observables
based on low-energy effective theories 
that replace Fermi liquid theory for the unconventional metals
\cite{
PhysRevB.8.2649,
PhysRevB.40.11571,
PhysRevB.50.14048,
PhysRevB.50.17917,
PhysRevLett.63.680,
polchinski1994low,
PhysRevB.46.5621,
nayak1994non,
PhysRevB.80.165102,
PhysRevB.82.075127,
PhysRevB.82.045121,
jiang2013non,
PhysRevB.88.245106,
PhysRevB.90.045121,
PhysRevB.92.041112}.

Antiferromagnetic (AF) quantum phase transitions arise in a wide range of layered compounds
\cite{PhysRevLett.105.247002,Hashimoto22062012,park2006hidden}.
Despite the recent progress made in field theoretic and numerical approaches to the AF quantum critical metal
\cite{
PhysRevLett.84.5608,
doi:10.1080/0001873021000057123,
PhysRevLett.93.255702,
PhysRevB.82.075128,
Abrahams28022012,
PhysRevB.91.125136,
PhysRevB.93.165114,
Berg21122012,
2015arXiv151207257S}, 
a full understanding of the non-Fermi liquid realized at the critical point has been elusive so far.
In two dimensions, strong quantum fluctuations and abundant low-energy particle-hole excitations 
render perturbative theories inapplicable.
What is needed is a non-perturbative approach 
which takes into account strong quantum fluctuations in a controlled way\cite{PhysRevB.90.045121}.

In this article, we present a non-perturbative field theoretic study 
of the AF quantum critical metal in two dimensions.
Although the theory becomes strongly coupled at low energies, 
we demonstrate that a small parameter which differs from the conventional coupling emerges dynamically.
This allows us to solve the strongly interacting theory reliably.
We predict the exact critical exponents 
that govern the scaling of dynamical and thermodynamic observables.

\section{Low-energy theory and interaction-driven scaling}

\begin{figure}[!ht]
\centering
 \includegraphics[scale=1.2]{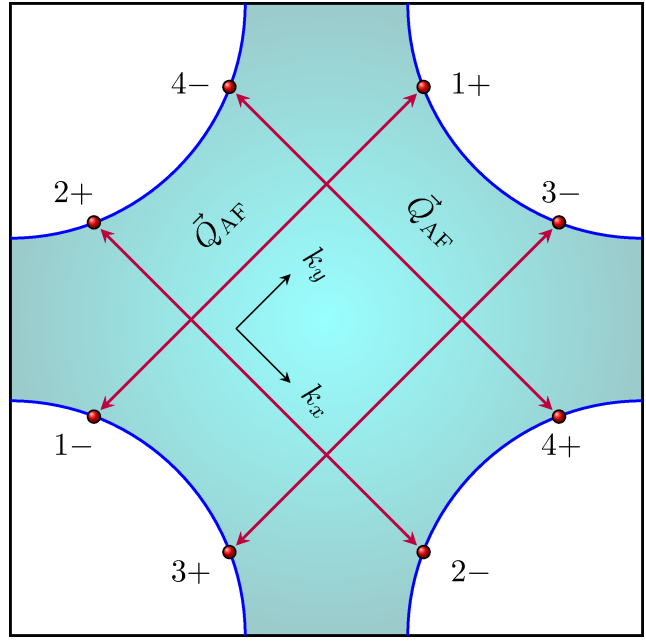}
 \caption{
A Fermi surface with the four-fold rotational symmetry. 
The (red) dots represent the hot spots connected by the AF wavevector $\vec{Q}_{AF}$. 
}
\label{fig:hot_spots}
\end{figure}

The relevant low-energy degrees of freedom at the metallic AF critical point 
are the AF collective mode and electrons near the hot spots, 
a set of points on the Fermi surface connected by the AF wavevector.
In the presence of the four-fold rotational symmetry 
and the reflection symmetry in two spatial dimensions,
there are generically eight hot spots, as is shown in  \fig{fig:hot_spots}.
Following Ref. \cite{PhysRevB.91.125136},
we write the action as 
\begin{align}
  \mathcal{S} &= 
\sum_{n=1}^4 
\sum_{\sigma=\uparrow, \downarrow}
\int 
d k~
\bar{\Psi}_{n,\sigma}(k) 
\Bigl[ i \gamma_0 k_0 + i \gamma_{1} \varepsilon_n(\vec{k}) \Bigr] 
\Psi_{n,\sigma}(k)
\nn 
&  + \frac{1}{4} 
\int d q ~
\Bigl[ q_0^2 + c_0^2 | \vec q |^2 \Bigr]
\tr { \Phi(-q)  ~ \Phi(q) } \nn
  & + i g
\sum_{n=1}^4
\sum_{\sigma,\sigma'}
\int 
d k d q ~
\Bigl[
\bar{\Psi}_{\bar n,\sigma}(k+q) 
\Phi_{\sigma,\sigma'}(q)
\gamma_{1} 
\Psi_{n,\sigma'} (k) 
\Bigr]  \nn
&  + u \int 
dk_1 d k_2  d q ~
   \tr { \Phi(k_1+q)  \Phi(k_2-q) } \tr { \Phi(- k_1)  \Phi(- k_2) }.
    \label{eq:3D_theory}
\end{align}
Here $k=(k_0,\vec k)$ denotes 
Matsubara frequency and two-dimensional momentum $\vec k =(k_x,k_y)$
with $dk \equiv \frac{d^3k}{(2 \pi)^3}$.
The four spinors are defined by
$\Psi_{1,\sigma}=(\psi^{(+)}_{1,\sigma},\psi^{(+)}_{3,\sigma})^T$,
$\Psi_{2,\sigma}=(\psi^{(+)}_{2,\sigma},\psi^{(+)}_{4,\sigma})^T$,
$\Psi_{3,\sigma}=(\psi^{(-)}_{1,\sigma}, - \psi^{(-)}_{3,\sigma})^T$ 
and
$\Psi_{4,\sigma}=(\psi^{(-)}_{2,\sigma}, - \psi^{(-)}_{4,\sigma})^T$,
where $\psi^{(m)}_{n,\sigma}$'s are electron fields 
with spin $\sigma=\uparrow, \downarrow$
near the hot spots labeled by $n=1,2,3,4$, $m=\pm$.
$\bar \Psi_{n,\sigma}= \Psi^\dagger_{n,\sigma} \gamma_0$,
where $\gamma_0 = \sigma_y, \gamma_1 = \sigma_x$
are $2 \times 2$ gamma matrices for the spinors.
The energy dispersions 
of the electrons near the hot spots are written as
$\varepsilon_1(\vec k) = v k_x + k_y$,
$\varepsilon_2(\vec k) = - k_x + v k_y$,
$\varepsilon_3(\vec k) = v k_x - k_y$ and
$\varepsilon_4(\vec k) = k_x + v k_y$,
where $\vec k$ represents the deviation of momentum away 
from each hot spot.
The commensurate AF wavevector ${\vec Q}_{AF}$ 
is chosen to be parallel to the $x$ and $y$ directions
modulo reciprocal lattice vectors.
The component of the Fermi velocity parallel to ${\vec Q}_{AF}$ 
at each hot spot is set to have unit magnitude.
$v$ measures the  component of the 
Fermi velocity perpendicular to ${\vec Q}_{AF}$.
$\Phi(q) = \sum_{a=1}^{3} \phi^a(q) \tau^a$ 
is a $2 \times 2$ matrix boson field
that represents the fluctuating AF order parameter,
where the $\tau^a$'s are the generators of the $SU(2)$ spin. 
$c_0$ is the velocity of the AF collective mode.
$g$ is the coupling between the collective mode
and the electrons near the hot spots.
$\bar n$ represents the hot spot connected to $n$ via $\vec Q_{AF}$ :
$\bar 1 = 3, \bar 2 = 4, \bar 3=1, \bar 4=2$.
$u$ is the  quartic 
coupling between the collective modes.

In two dimensions, the conventional perturbative expansion becomes unreliable 
as the couplings grow at low energies.
Since the interaction plays a dominant role,
we need to include the interaction
up front rather than treating it
as a perturbation to the kinetic energy. 
Therefore, we start with an interaction-driven scaling\cite{PhysRevB.90.045121}
in which the fermion-boson coupling is deemed marginal.
Under such a scaling, one cannot keep all the kinetic terms as marginal operators.
Here we choose a scaling that keeps
the fermion kinetic term marginal 
at the expense of making the boson kinetic term irrelevant. 
This choice will be justified through explicit calculations.
It reflects the fact that the dynamics of the boson is dominated by 
particle-hole excitations near the Fermi surface in the low-energy limit,
unless the number of bosons per fermion is infinite\cite{PhysRevB.88.125116}.
The marginality of the fermion kinetic term and the fermion-boson coupling 
uniquely fixes the dimensions of momentum and the fields 
under the interaction-driven tree-level scaling, 
\bqa
&& [k_0] = [k_x] = [k_y] = 1, \nn
&& [\psi(k)] = [\phi(k)] = -2.
\label{AN1}
\eqa
Under this scaling, the electron keeps the classical scaling dimension,
while the boson has an $O(1)$ anomalous dimension compared to the Gaussian scaling.
At this point, \eq{AN1} is merely an Ansatz.
The real test is to show that these exponents are actually exact, 
which is the main goal of this paper.

Under \eq{AN1}, the entire boson kinetic term and the quartic coupling are irrelevant.
The minimal action which includes only marginal terms is written as
\begin{align}
  \mathcal{S} &= 
\sum_{n=1}^4 
\sum_{\sigma=\uparrow, \downarrow}
\int 
d k~
\bar{\Psi}_{n,\sigma}(k) 
\Bigl[ i \gamma_0 k_0 + i \gamma_{1} \varepsilon_n(\vec{k}) \Bigr] 
\Psi_{n,\sigma}(k)
\nn 
  & + i  \sqrt{ \frac{\pi v}{2} }  
  \sum_{n=1}^4
\sum_{\sigma,\sigma'}
\int 
d k d q ~
\Bigl[
\bar{\Psi}_{\bar n,\sigma}(k+q) 
\Phi_{\sigma,\sigma'}(q)
\gamma_{1} 
\Psi_{n,\sigma'} (k) 
\Bigr].
\label{eq:min_theory}
\end{align}
Here, the fermion-boson coupling is set to be proportional to $\sqrt{v}$ by rescaling the boson field.
The Yukawa coupling is replaced with $\sqrt{v}$
because the interaction is screened
such that $g^2$ becomes $O(v)$ 
in the low-energy limit\cite{PhysRevB.91.125136}.
Although $g$ and $v$ can be independently tuned in the microscopic theory,
they rapidly flow to a universal line defined by $g^2 \sim v$ at low energies\cite{2017arXiv170108218L}.
\eq{eq:min_theory} should be understood as the minimal theory
that captures the universal physics at low energies,
where the dynamics of the collective mode 
is dominated by particle-hole excitations
rather than the bare kinetic term,
and $v$ is the only dimensionless parameter.
In the small $v$ limit, $g$ also vanishes  
because a nested Fermi surface 
provides a large phase space for 
low-energy particle-hole excitations with momentum $\vec Q_{AF}$
that screen the interaction. 
Even when $g,v$ are small,
this is a strongly interacting theory
because $g^2/v \sim 1$  is the expansion parameter
in the conventional perturbative series.
With $g^2/v \sim 1$, 
the leading boson kinetic term 
which is generated from particle-hole excitations is $O(1)$, 
as will be seen later.

\begin{figure}[!ht]
\centering
 \includegraphics[scale=1.5]{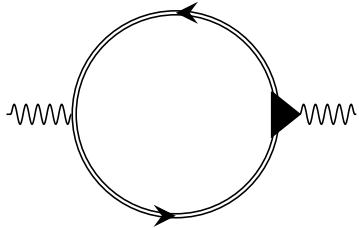}
 \caption{
The exact boson self-energy.
The double line is the fully dressed fermion propagator.
The triangle represents the fully dressed vertex.
}
\label{fig:SD}
\end{figure}

\section{Self-consistent solution}

Naively the theory is singular due to the absence of a boson kinetic term.
However, particle-hole excitations generate a self-energy
which provides non-trivial dynamics for the collective mode.
The Schwinger-Dyson equation for the boson propagator (shown in \fig{fig:SD}) reads 
\bqa
D(q)^{-1} &=& 
m_{CT}
-  \pi v \sum_n
\int dk ~ 
\mbox{Tr}\lt[ \gamma_{1} G_{\bar n}(k + q) 
\Gamma(k,q)  G_n(k) \rt] .
\label{eq:SD}
\eqa
Here $D(k)$, $G(k)$ and $\Gamma(k,q)$ represent 
the fully dressed propagators of the boson and the fermion,
and the vertex function, respectively.
$m_{CT}$ is a mass counter term that is added 
to tune the renormalized mass to zero. 
The trace in \eq{eq:SD} is over the spinor indices.
It is difficult to solve the full self-consistent equation 
because $G(k)$ and $\Gamma(k,q)$ depend on the unknown $D(q)$.
One may use $v$ as a small parameter to solve the equation.
The one-loop analysis shows that $v$ flows to zero 
due to emergent nesting of the Fermi surface near the hot spots\cite{PhysRevLett.84.5608,
doi:10.1080/0001873021000057123,
PhysRevB.82.075128,
PhysRevB.93.165114}.
This has been also confirmed in the $\epsilon$ expansion
based on the dimensional regularization scheme\cite{PhysRevB.91.125136,2017arXiv170108218L}.
Of course, the perturbative result valid close to three dimensions 
does not necessarily extend to two dimensions.
Nonetheless, we show that this is indeed the case.
Here we proceed with the following steps:
\begin{enumerate}
\item we solve the Schwinger-Dyson equation for the boson propagator in the small $v$ limit, 
\item we show that  $v$ flows to zero at low energies by using the boson propagator obtained under the assumption of $v \ll 1$. 
\end{enumerate}
We emphasize that the expansion in $v$ is different from the conventional perturbative expansion in coupling.
Rather it involves a non-perturbative summation over an infinite series of diagrams as will be shown in the following.

We discuss step 1) first.
In the small $v$ limit, 
the solution to the Schwinger-Dyson equation is 
\bqa
D(q)^{-1} = |q_0| + c(v) \Big[ |q_x| + |q_y| \Big],
\label{eq:D}
\eqa
where the `velocity' of the strongly damped collective mode is given by
\bqa
c(v) = \frac{1}{4} \sqrt{  v \log (1/v) }.
\label{eq:cv}
\eqa
Solving the Schwinger-Dyson equation consists of two parts.
First, we assume \eq{eq:D} 
with a hierarchy of the velocities $v \ll c(v) \ll 1$
as an Ansatz
to show that only the one-loop vertex correction is important in \eq{eq:SD}.
Then we show that Eqs. (\ref{eq:D}) and (\ref{eq:cv}) actually satisfy
\eq{eq:SD} with the one-loop dressed vertex.

We begin by estimating the magnitude of general diagrams,
assuming that the fully dressed boson propagator is given by \eq{eq:D} with \eq{eq:cv} 
in the small $v$ limit.
In general, the integrations over loop momenta
diverge in the small $v$ limit as fermions and bosons lose their dispersion in some directions.
In each fermion loop, the component of the internal momentum tangential to the Fermi surface 
is unbounded in the small $v$ limit due to nesting.
For a small but nonzero $v$, the divergence is cut off at a scale proportional to $1/v$, 
and each fermion loop contributes a factor of $1/v$.
Each of the remaining loops necessarily has at least one boson propagator.
For those loops, the momentum along the Fermi surface is cut off by the energy of the boson 
which provides a lower cut-off momentum proportional to $1/c$ for $c \gg v$.
Therefore, the magnitude of a general $L$-loop diagram with $V$ vertices, $L_f$ fermion loops 
and $E$ external legs is at most 
\bqa
I \sim v^{ V/2 - L_f} c^{-(L-L_f)} \sim v^{ \frac{1}{2}(E-2)} \left( \frac{v}{c} \right)^{(L -L_f) },
\label{EST}
\eqa
where $V = 2L + E-2$ is used.
Higher-loop diagrams are systematically suppressed with increasing  $(L-L_f)$ provided $v \ll c$.
This is analogous to the situation where a ratio between velocities is 
used as a control parameter in a Dirac semi-metal\cite{PhysRevB.78.064512}
\footnote{
There also has been an attempt to use a different ratio of velocities
as a control parameter in non-Fermi liquids
with critical bosons centered at zero momentum
[A. Fitzpatrick, S. Kachru, J. Kaplan, S. A. Kivelson, S. Raghu, arXiv:1402.5413].
}.
 If \eq{eq:cv} holds, the upper bound becomes 
$I \sim v^{ \frac{1}{2}(E-2) + \frac{1}{2}(L -L_f) }$ up to a logarithmic correction.
It is noted that \eq{EST} is only an upper bound 
because some loop integrals which involve un-nested fermions 
remain finite even in the small $v$ limit. 
Some diagrams can also be smaller than the upper bound
because their dependences on external momentum 
are suppressed in the small $v$ and $c$ limit.
A systematic proof of \eq{EST}  is available 
in Appendix A.

\begin{figure}[!ht]
\centering
\subfigure[]
{\includegraphics[scale=1.5]{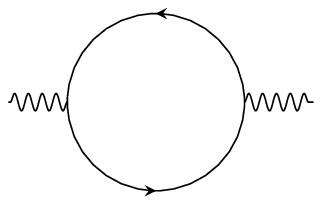}
\label{fig:SEb}} ~~~~~
\subfigure[]
{\includegraphics[scale=1.5]{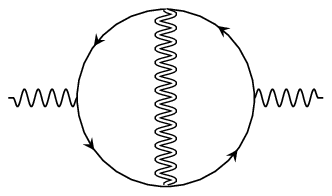}
\label{fig:2LSEb}} ~~~~~
\caption{
The leading order diagrams for the boson self-energy
in the small $v$ limit.   
Solid lines are the bare fermion propagators.
The wiggly double line represents the boson propagator 
 consistently dressed with the self-energy in (a) and (b).
The dressed boson propagator includes an infinite series 
of nested self-energies with a fractal structure.
}
\label{fig: CT1}
\end{figure}

For $v \ll c$,
the leading order contribution 
for the boson self-energy ($E=2$)
is generated from \fig{fig:SEb},
which is the only diagram that satisfies $L=L_f$.
All other diagrams are sub-leading in $v$.
However, this is not enough 
because the one-loop diagram gives 
$D(q)^{-1} = |q_0|$, which  is independent of spatial momentum. 
One has to include the next order diagram  (\fig{fig:2LSEb}) 
which generates a dispersion.
Therefore, \eq{eq:SD} is reduced to  
\bqa
&& D(q)^{-1}  = 
m_{CT}^{'}
+  |q_0|  \nn
&& - \frac{ \pi^2 v^2}{2} \sum_n \int dp ~ dk ~
\tr{\gamma_{1} G_{n}^{(0)}(k+p) \gamma_{1}  G_{\bar n}^{(0)}(p+q+k) \gamma_{1}  G_{ n}^{(0)}(q+k) \gamma_{1} G_{\bar n}^{(0)}(k)} D(p).  \nn
\label{eq:SD2}
\eqa
Here $m_{CT}^{'}$ is a two-loop mass counter term.
We can use the free fermion propagator $G^{(0)}_n$
because the fermion self-energy correction 
is sub-leading in $v$. 
An explicit calculation 
of \eq{eq:SD2} confirms that 
the self-consistent boson propagator 
takes the form of \eq{eq:D}.
The boson velocity satisfies
the self-consistent equation
$c = \frac{v}{8c} \log (c/v)$,
which is solved by \eq{eq:cv}
in the small $v$ limit.
$c$ is much larger than $v$ in the small $v$ limit 
because of the enhancement factor $1/c$ in the two-loop diagram :
the collective mode speeds up itself 
through enhanced quantum fluctuations
if it gets too slow.
We note that the anti-screening nature
of the vertex correction associated with the non-Abelian $SU(2)$ vertex, 
$\sum_{a=1}^3 \tau^a \tau^b \tau^a = - \tau^b$,
is crucial to generate the right sign for the 
boson kinetic term\cite{PhysRevB.94.195135}.
This does not hold for 
Ising-like or XY-like spin fluctuations\cite{PhysRevLett.115.186405}.
The details on the computation of \eq{eq:SD2} are available in Appendix B.
It is noted that \eq{eq:SD2}
constitutes a non-perturbative sum over an infinite series of diagrams
beyond the random phase approximation (RPA).
The dynamics of the boson 
generated from the fermionic sector
dominates at low energies.
This justifies the choice to drop the bare 
kinetic term in \eq{eq:min_theory}.

So far, we have assumed that $v$ is small 
to obtain the self-consistent dynamics of the AF collective mode.
Now we turn to step 2) and show that $v$ indeed flows to zero in the low-energy limit.
According to \eq{EST},
the leading quantum corrections to the local action in \eq{eq:min_theory}
are the one-loop diagrams for the fermion self-energy and the vertex function.
However, the momentum-dependent one-loop fermion self-energy 
happens to be smaller than what is expected from \eq{EST} by an additional power of $c \sim \sqrt{v}$.
This is because the dependence on the external momentum is suppressed in the small $c$ limit for the one-loop self-energy.
As a result, we include the fermion self-energy up to two loops 
in order to capture all quantum corrections to the leading order in $v$.
All other higher-loop diagrams are negligible in the small $v$ limit.
The self-energy and vertex correction are logarithmically divergent in a UV cut-off.
Counter terms are added such that the renormalized quantum effective action
becomes independent of the UV cut-off.
The full details on the computation of the counter terms and the  beta function
can be found in Appendix C.
The bare action that includes the counter terms is obtained to be
\begin{align}
  \mathcal{S}_B &= 
\sum_{n=1}^4 
\sum_{\sigma=\uparrow, \downarrow}
\int 
d^3 k~
\bar{\Psi}_{n,\sigma}(k) 
\Bigl[ i Z_1 \gamma_0 k_0    + i \gamma_{1} \varepsilon_n^B(\vec{k}) \Bigr] 
\Psi_{n,\sigma}(k)
\nn 
 & + i Z_6  \sqrt{ \frac{\pi v}{2} }    \sum_{n=1}^4
\sum_{\sigma,\sigma'}
\int  d^3 k d^3 q ~
\Bigl[
\bar{\Psi}_{\bar n,\sigma}(k+q) 
\Phi_{\sigma,\sigma'}(q)
\gamma_{1} 
\Psi_{n,\sigma'} (k) 
\Bigr],
\label{eq:bare_theory}
\end{align}
where 
$\varepsilon_1^B(\vec k) = Z_2 v k_x + Z_3 k_y$,
$\varepsilon_2^B(\vec k) = - Z_3 k_x + Z_2 v k_y$,
$\varepsilon_3^B(\vec k) = Z_2 v k_x - Z_3 k_y$ and
$\varepsilon_4^B(\vec k) = Z_3 k_x + Z_2 v k_y$
with
$Z_1 = 1 - \f{3}{4 \pi} \f{v}{c(v)} \log\lt( \f{\Lambda}{\mu} \rt)$,
$Z_2 = 1 + \f{15}{4 \pi^2} v \, \log\lt(\f1{c(v)} \rt) \, \log\lt( \f{\Lambda}{\mu} \rt)$,
$Z_3 = 1 - \f{9}{4 \pi^2} v \, \log\lt(\f1{c(v)} \rt) \, \log\lt( \f{\Lambda}{\mu} \rt)$ and
$Z_6 = 1 - \f{1}{4 \pi} \f{v}{c(v)} \log\lt( \f{c(v)}{v} \rt) \log\lt( \f{\Lambda}{\mu} \rt)$.
Here $\Lambda$ is a UV cut-off
above which non-linear terms 
in the fermionic dispersion
become important. 
$\mu$ is the scale at which the physical
propagators and vertex function are expressed  
in terms of $v$ through the renormalization conditions,
$\frac{-i}{2}\left. \frac{\partial}{\partial k_0 } \tr{ \gamma_0 G_{1}(k)^{-1} } \right|_{k=(\mu, 0,0)} = 1 + F_1(v)$,
$ \frac{-i}{2}\left. \frac{\partial}{\partial k_x } \tr{ \gamma_1 G_{1}(k)^{-1} } \right|_{k=(0, \mu,0)}= v \left( 1 + F_2(v) \right)$,
$\frac{-i}{2} \left. \frac{\partial}{\partial k_y } \tr{ \gamma_1 G_{1}(k)^{-1} } \right|_{k=(0,0,\mu)}= 1 + F_3(v)$,
$\frac{1}{2} \left. \tr{ \gamma_1 \Gamma(k,q) } \right|_{q=0,k=(\mu,0,0)}=   1 + F_4(v)$,
where the $F_i(v)$'s are UV-finite functions of $v$, 
which vanish in the small $v$ limit.
The specific form of $F_i(v)$ is unimportant, 
and they can be changed by adding finite counter terms in $Z_i$. 
$G_n(k)$ with $n=2,3,4$ are fixed from $G_1(k)$ by the four-fold rotational symmetry.
The bare and renormalized variables are related to each other through
$k_{B,x} =k_{x}$,
$k_{B,y} =k_{y}$,
$k_{B,0} =\frac{Z_1}{Z_3} k_0$,
$v_{B} = \frac{Z_2}{Z_3}v$,
$\Psi_{B}(k_B) = \frac{Z_3}{Z_1^{\f{1}{2}}}\Psi(k)$,
$\Phi_{B}(k_B) = \f{Z_3^{\f12} Z_6}{Z_1 Z_2^{\f12}} \, \Phi(k)$.
By requiring that the bare quantities are independent of $\mu$,
we obtain the beta function $\beta_v \equiv  \frac{dv}{d\log \mu}$,
which dictates the dependence of the renormalized velocity on the scale,
\bqa
\beta_v =  \f{6}{\pi^2} \, v^{2} \log\lt( \f1{c(v)} \rt).
\label{eq:beta2}
\eqa
As a function of the energy scale $\mu$, 
$v$ is renormalized according to
\bqa
\frac{ d v}{d \log \mu} =  \f{6}{\pi^2} \, v^{2} \log\lt( \f1{c(v)} \rt).
\label{eq:beta}
\eqa
If $v$ is initially small, 
\eq{eq:beta} is reliable.
It predicts that $v$ becomes even smaller and flows to zero as 
\bqa
v = \frac{\pi^2}{3} \left( \log \frac{1}{\mu} ~ \log \log \frac{1}{\mu} \right)^{-1}
\label{eq:v}
\eqa 
in the small $\mu$ limit.
The way $v$ flows to zero in the low-energy limit does not depend on the initial value of $v$. 
This completes the cycle of self-consistency.
\eq{eq:D} obtained in the small $v$ limit becomes asymptotically 
exact in the low-energy limit
within a nonzero basin of attraction in the space of $v$ whose fixed point is $v=0$.
The dynamical critical exponent and the anomalous dimensions are given by
\bqa
	z &=&   1 + \f{3}{4 \pi} \f{v}{c(v)}, \nn
	\eta_\phi &=&  \f{1}{4\pi} \f{v}{c(v)} \log\lt( \f{c(v)}{v} \rt), \nn
	\eta_\psi &= &  - \f{3}{8 \pi} \f{v}{c(v)}
\label{eq:exponents}
\eqa
to the leading order in $v$.
Here $z$ sets the dimension of frequency relative to momentum. 
$\eta_\phi$, $\eta_\psi$ are the corrections 
to the  interaction-driven tree-level scaling dimensions of the boson and fermion, respectively.
The critical exponents are controlled by $w \equiv v/c(v)$,
which flows to zero as 
$w = \frac{4 \pi}{\sqrt{3}} \left( \log^{1/2} \frac{1}{\mu} ~ \log \log \frac{1}{\mu} \right)^{-1}$ 
in the low-energy limit.
This confirms that the scaling dimensions in \eq{AN1} become asymptotically exact in the low-energy limit.
This is compatible with the fact that
an inclusion of higher-loop corrections in the $\epsilon$-expansion 
reproduces $z=1$, irrespective of $\epsilon$\cite{2017arXiv170108218L}.

\section{Physical observables}

Although $z-1$, $\eta_\psi$ and $\eta_\phi$ vanish in the low-energy limit,
the sub-logarithmic decay of $w$ with energy introduces corrections to the correlation functions at intermediate energy scales,
which are weaker than power-law but stronger than logarithmic corrections\cite{varma1989phenomenology}.
The retarded Green's function for the hot spot $1+$ takes the form,
\bqa
G^R_{1+}(\omega, \vec k) =
\frac{1}{ 
F_\psi( \omega )
\left[
\omega ~ F_z(\omega)  
\left(
1 + i \frac{  \sqrt{3} \pi }{2}   \frac{1}{ \sqrt{ \log \frac{1}{\omega} } ~ \log \log \frac{1}{\omega}  }
\right)
- \left(
\frac{ \pi^2 }{3}  \frac{k_x}{  \log \frac{1}{\omega} ~ \log \log \frac{1}{\omega}  }
+ k_y
\right)
\right]
}
\label{eq:GR}
\eqa
in the small $\omega$ limit with the ratio $\frac{ \vec k}{ \omega ~ F_z(\omega)  }$ fixed. 
Here $\omega$ is the real frequency.
$F_\psi(\omega)$ and $F_z(\omega)$ 
are  functions which capture the contributions 
from $\eta_\psi$ and $z$ at intermediate energy scales.  
In the small $\omega$ limit, they are given by
\bqa
F_\psi(\omega) = \left( \log \frac{1}{\omega}  \right)^{\frac{3}{8} }, ~~~
F_z(\omega) = e^{ 2 \sqrt{3} \frac{ \left( \log \frac{1}{\omega}  \right)^{1/2}}{ \log \log \frac{1}{\omega} }}.
\eqa
$F_\psi$ and $F_z$ only contribute as sub-leading corrections
instead of modifying the exponents.
However, they are still parts of the universal data 
that characterizes the critical point\cite{PhysRevB.82.075128}.
The additional logarithmic suppression in the dependence of $k_x$ 
is due to $v$ which flows to zero in the low-energy limit.
The local shape of the Fermi surface is deformed as
$k_y \sim \frac{k_x}{\log 1/k_x \log \log 1/k_x}$.
The scaling form of the Green's function at different hot spots
can be obtained by applying a sequence of $90$ degree rotations 
and a space inversion to \eq{eq:GR}.
The spectral function at the hot spots exhibits 
a power-law decay with the super-logarithmic correction as a function of frequency,
$
A(\omega) \sim 
\frac{1}{
\omega    F_z(\omega) F_\psi(\omega)    ( \log 1/ \omega )^{1/2} \log \log 1/\omega      
 }
$.

The retarded spin-spin correlation function is given by
\bqa
D^R(\omega, \vec q) =  \frac{1}{
F_{\phi} (\omega)
\left( 
 -i \omega F_z(\omega)
+
\frac{\pi }{4 \sqrt{3}}
\frac{|q_x| + |q_y|}{ \left( \log \frac{1}{\omega} \right)^{1/2}  }
\right)
}
\label{eq:DR}
\eqa
in the small $\omega$ limit
with fixed $\frac{ \vec q}{ \omega ~ F_z(\omega)  }$.
$F_{\phi}(\omega)$ is another universal function 
that describes the super-logarithmic correction of $\eta_\phi$,
\bqa
F_{\phi} (\omega)
= e^{ \frac{2 }{\sqrt{3}} \left( \log \frac{1}{ \omega }  \right)^{1/2}  }
\eqa
in the small $\omega$ limit.
The factor of $  \left( \log \frac{1}{\omega} \right)^{-1/2} $
in the momentum-dependent term is due to the boson velocity 
which flows to zero in the low-energy limit.
Due to the strong Landau damping,
the spin fluctuation is highly incoherent.
It will be of great interest to test the scaling forms in 
Eqs. (\ref{eq:GR}) and (\ref{eq:DR})  
from angle resolved photoemission spectroscopy 
and neutron scattering, respectively.

Now we turn to thermodynamic properties.
The total free energy density can be written as
$f = \frac{1}{2} ~Tr \left[   \log D^{-1} - \Pi D  \right]
-  Tr \left[  \log G^{-1}  - \Sigma G  \right]
+ \Phi_2$,
where $\Pi$, $\Sigma$ are the self-energies
of the boson and fermion respectively,
and $\Phi_2$ includes the two particle irreducible diagrams\cite{PhysRev.118.1417}.
Here, the traces sum over three momenta and flavors.
To the leading order in $v$, 
$f_B = \frac{1}{2} ~ Tr[   \log D^{-1} ]$
and $f_F =   Tr [  \log G^{(0)} ]$ dominate.
The dominant fermionic contribution comes from electrons away from the hot spots, 
$f_{F} \sim k_F T^2$,
where $k_F$ is the size of the Fermi surface.
Naively, the bosonic contribution is expected to obey hyperscaling,
because low-energy excitations are confined near the ordering vector.
However, the free energy of the mode with momentum $\vec p$  is
suppressed only algebraically as $\frac{T^2}{c ( |p_x|+ |p_y| )}$ at large momenta,
in contrast to the exponential suppression for the free boson.
The slow decay is due to the incoherent nature of 
the damped AF spin fluctuations,
which have a significant spectral weight at low energies even at large momenta.
As a result, $f_B \sim \int d\vec p ~ \frac{T^2}{c ( |p_x|+ |p_y| )} $ 
is UV divergent.
In the presence of the irrelevant local kinetic term,
$\frac{c_0^2}{\td \Lambda} |\vec p|^2$
with $c_0 \sim 1$,
the momentum integration is cut-off at
$p_{max} \sim c \td \Lambda$,
and $f_B$ is proportional to $\td \Lambda$.
From the scaling equation for $f_B$,
$\left[
zT \frac{\partial}{\partial T}
+ \td \Lambda \frac{\partial}{\partial \td \Lambda}
- \beta_c \frac{ \partial}{\partial c}
- (2+z)
\right] f_{B}(T,c,\td \Lambda) = 0$,
we obtain
$ f_B \sim \td \Lambda T^2 F_z(T)$
in the low temperature limit.
Remarkably, the bosonic contribution violates the hyperscaling,
and it is larger than the fermionic contribution at low temperatures.
In this case, the power-law violation of the hyperscaling 
is a consequence of the $z=1$ scaling
rather than the fact that $v,c$ flow to zero\cite{PhysRevB.92.165105}.
The free energy gives rise to the specific heat 
which exhibits the $T$-linear behavior with the super-logarithmic correction, 
\bqa
c_V \sim \tilde \Lambda T F_z (T).
\label{eq:c}
\eqa  
It is noted the deviation from the $T$-linear behavior 
is stronger than a simple logarithmic correction
because $F_z(T)$ includes all powers of $\sqrt{ \log \frac{1}{T}}$.
%


If the system is tuned away from the critical point,
the boson acquires a mass term, $(\lambda-\lambda_c) \int dq \tr{ \Phi_q \Phi_{-q}}$,
where $\lambda$ is a tuning parameter. 
Due to the suppression of higher-loop diagrams, 
the scaling dimension of $\Phi^2$  is $-4$ in momentum space.
This implies that $\nu=1$ in the low-energy limit, 
which is different from the mean-field exponent.
The power-law scaling of the correlation length $\xi$ with $\lambda$
is modified by a super-logarithmic correction,
\bqa
\xi \sim ( \lambda - \lambda_c)^{-1} 
F_{\xi}(  \lambda - \lambda_c),
\eqa
where $F_{\xi}( \delta \lambda )$ is a universal function which embodies
both the anomalous dimension of the boson 
and the vertex correction for the mass insertion.
The former dominates close to the critical point,
and $F_{\xi}( \delta \lambda )$ is the same as $F_{\phi}(  \delta \lambda)$ 
to the leading order in small $\delta \lambda$.
The derivation of the scaling forms of the physical observables is available 
in Appendix D.


The scaling forms of the physical observables
discussed above are valid in the low energy limit.
At high energies, 
there will be crossovers to different behaviors.
The first crossover is set by the scale below which
the dynamics of the collective mode 
is dominated by particle-hole excitations,
and therefore Eqs. (\ref{eq:DR}) and (\ref{eq:c}) hold.
It is determined by the competition between 
\eq{eq:D} and the irrelevant local kinetic term for the collective mode in \eq{eq:3D_theory}.
For $\omega < \frac{c(v)^2}{c_0^2} \tilde \Lambda$,
the terms linear in frequency and momentum dominate,
where $\tilde \Lambda$ is an energy scale associated with the irrelevant kinetic term.
The details on the crossover are described in Appendix B.
In the small $v$ limit with $c_0 \sim 1$,
this crossover scale for the boson goes as $E_b^* \sim c^2 \tilde \Lambda$.
The second crossover scale, denoted as $E_f^*$, 
is the one below which 
the behavior of the fermions at the hot spots 
deviates from the Fermi liquid one.
For a small but non-zero $v$, 
the leading order self-energy correction to the fermion propagator
is $\frac{3}{4\pi} \frac{v}{c(v)} \omega \log \frac{\Lambda}{\omega}$,
which becomes larger than the bare term
for $\omega < E_f^*$ with 
$E_f^* \sim \Lambda e^{-\frac{\pi}{3}\sqrt{ \frac{\log1/v}{v} }}$.
Since $v$ flows to zero only logarithmically, 
the flow of $v$ can be ignored
for the estimation of $E_f^*$.
The value of $v$ changes appreciably
below $\Lambda e^{-\frac{1}{v \log 1/v}}$
as is shown in Appendix C.

At sufficiently low temperatures, 
the system eventually becomes unstable against pairing.
An important question is how the crossover scales compare 
with the superconducting transition temperature $T_c$.
The spin fluctuations renormalize pairing interactions
between electrons near the hot spots, and  
enhance  $d$-wave superconductivity\cite{scalapino1986d,PhysRevB.34.6554,doi:10.1143/JPSJ.59.2905,Berg21122012,2015arXiv151204541L}.
In the small $v$ limit, however,
the renormalization of the pairing interaction 
by the AF spin fluctuations is suppressed by $\frac{v}{c(v)}$
for the same reason that the vertex correction is suppressed.
Because the Yukawa coupling is marginal at the fixed point,
it adds an additional logarithmic divergence to the 
usual logarithmic divergence caused by the BCS instability\cite{PhysRevD.59.094019,PhysRevB.46.14803,PhysRevB.72.174520}.
The pairing vertex is enhanced by
$\alpha \frac{v}{c} \log \frac{\Lambda}{\omega} \log \frac{ E_b^* }{\omega}$ 
with $\alpha \sim 1$ at frequency $\omega$. 
The first logarithm is from the usual BCS mechanism.
The second logarithm is from the gapless spin fluctuations,
where $E_b^* \sim c^2 \tilde \Lambda$ is the energy cut-off for the spin fluctuations
in the small $c$ limit
as is shown in Appendix B. 
This gives $T_c \sim  c \sqrt{ \Lambda  \tilde \Lambda} e^{-   \sqrt{ \frac{c}{\alpha ~ v} } }$.
Although $T_c$ is enhanced by the critical spin fluctuations, 
it remains exponentially small in
$\sqrt{ \frac{c(v)}{v} } \sim v^{-\frac{1}{4}}$ 
in the small $v$ limit.
There is a hierarchy among the energy scales,
$E_f^* \ll
T_c 
\ll E_b^* $
in the small $v$ limit.
This suggests that 
the system undergoes a superconducting transition
before the fermions at the hot spots lose coherence.
On the one hand, this is similar to the nematic quantum critical point in two dimensions
where the system is prone to develop a superconducting instability
before the coherence of quasiparticles breaks down\cite{PhysRevB.91.115111,PhysRevLett.114.097001}.
On the other hand, even without superconductivity, 
the fermions  are only weakly perturbed by the spin fluctuations
in the present case. 
It is the collective mode that is heavily dressed by quantum effects.
For the collective mode,
there is a large window between $T_c$ and $E_b^*$
within which the universal scaling given by \eq{eq:D} is obeyed.
The size of the energy window for the critical scaling 
is non-universal due to the slow flow of $v$, 
and it depends on the bare value of $v$.
Our prediction is that 
there is a better chance to observe the $z=1$ critical scaling above $T_c$,
and the enhancement of $T_c$ by AF spin fluctuations
is rather minimal\cite{Horio2016}
in materials whose bare Fermi surfaces are closer to  perfect nesting near the hot spots.

\section{Summary and Discussion}

In summary, we solve the low-energy field theory
that describes the antiferromagnetic quantum critical metal
in two spatial dimensions.
We predict the exact critical exponents
which govern the universal scaling of physical observables
at low temperatures.
Finally, we comment on 
earlier theoretical approaches,
and provide a comparison with experiments.


Our results are qualitatively different
from earlier theoretical works
\cite{
PhysRevLett.84.5608,
doi:10.1080/0001873021000057123,
PhysRevLett.93.255702,
PhysRevB.82.075128,
Abrahams28022012,
PhysRevB.93.165114}
which have invariably predicted
the dynamical critical exponent $z$
to be larger than one.
In particular, if one uses the one-loop dressed propagators with $z=2$, 
individual higher-loop corrections are logarithmically divergent at most.
However, this does not imply that the higher-loop corrections are small.
The logarithmic corrections remain important in two dimensions 
due to the strong coupling nature of the theory,
and they can introduce $\mathcal O(1)$ anomalous dimensions. 
The one-loop analysis based on the dimensional regularization scheme
also predicts that the dynamical critical exponent is $z= 1 + \mathcal O(\epsilon)$
in $3-\epsilon$ space dimensions\cite{PhysRevB.91.125136}.
It turns out that it is not enough to include 
only the one-loop corrections 
even to the leading order in $\epsilon$
due to an infrared singularity associated 
with the emergent quasi-locality\cite{PhysRevB.94.195135}.
Once all quantum corrections are taken into account 
to the leading order in $\epsilon$ consistently,
the dynamical critical exponent becomes $z=1$ again\cite{2017arXiv170108218L}
in agreement with the current result.
The key that makes the present theory solvable 
is the emergent hierarchy of the velocities $v \ll c(v)$,
which becomes manifest only after 
quantum fluctuations are included consistently\cite{2017arXiv170108218L}.


Now we make an attempt to compare our predictions with experiments.
Electron doped cuprates are probably the simplest examples
of quasi-two-dimensional compounds 
that exhibit antiferromagnetic phase transitions
in the presence of itinerant electrons, 
without having extra degrees of freedom such as local moments 
or extra bands.
In the normal state of the optimally doped $\mathrm{Pr_{0.88} La Ce_{0.12} Cu O_{4-\delta}}$,
inelastic neutron scattering
shows an overdamped AF spin fluctuation peaked at $(\pi,\pi)$
whose width in  momentum space exhibits 
a weak growth with increasing energy\cite{Wilson2006}.
The theoretical prediction from \eq{eq:DR}
is that the width of the incoherent peak
scales linearly with energy 
upto a super-logarithmic correction
in the low energy limit.
However, it is hard to make a quantitative comparison 
due to the limited momentum resolution in the experiment.
In $\mathrm{Nd_{2-x} Ce_x Cu O_{4\pm \delta}}$ (NCCO), 
 inelastic neutron scattering suggests
that the magnetic correlation length $\xi$
scales inversely with temperature near the critical doping\cite{Motoyama2007}.
Furthermore, $\xi$ measured at the pseudogap temperature
diverges as $(x-x_c)^{-1}$.
If interpreted in terms of the clean AF quantum critical scenario,
which may be questionable due to disorder,
this is consistent with $z=1$ and $\nu=1$.
Angle resolved photoemission spectroscopy (ARPES) for NCCO shows  
a reduced quasiparticle weight at the hot spots\cite{PhysRevLett.87.147003,PhysRevB.78.100505}.
This is in qualitative agreement with 
the prediction of \eq{eq:GR},
which implies that the quasiparticle weight vanishes at the hot spots,
as compared to the region away from the hot spots
where quasiparticles are well defined.
Although the spectroscopic measurements 
are in qualitative agreement with the theoretical predictions,
we believe that more experiments are needed
to make quantitative comparisons.
On the theoretical side, 
transport properties need to be better understood,
for which electrons away from hot spots are expected to 
play an important role.

\bibliography{sdw}

\begin{thebibliography}{55}
\expandafter\ifx\csname natexlab\endcsname\relax\def\natexlab#1{#1}\fi
\expandafter\ifx\csname bibnamefont\endcsname\relax
  \def\bibnamefont#1{#1}\fi
\expandafter\ifx\csname bibfnamefont\endcsname\relax
  \def\bibfnamefont#1{#1}\fi
\expandafter\ifx\csname citenamefont\endcsname\relax
  \def\citenamefont#1{#1}\fi
\expandafter\ifx\csname url\endcsname\relax
  \def\url#1{\texttt{#1}}\fi
\expandafter\ifx\csname urlprefix\endcsname\relax\def\urlprefix{URL }\fi
\providecommand{\bibinfo}[2]{#2}
\providecommand{\eprint}[2][]{\url{#2}}

\bibitem[{\citenamefont{Landau}(1957)}]{LFL}
\bibinfo{author}{\bibfnamefont{L.}~\bibnamefont{Landau}},
  \bibinfo{journal}{Sov. Phys. JETP} \textbf{\bibinfo{volume}{3}},
  \bibinfo{pages}{920} (\bibinfo{year}{1957}).

\bibitem[{\citenamefont{Hertz}(1976)}]{PhysRevB.14.1165}
\bibinfo{author}{\bibfnamefont{J.~A.} \bibnamefont{Hertz}},
  \bibinfo{journal}{Phys. Rev. B} \textbf{\bibinfo{volume}{14}},
  \bibinfo{pages}{1165} (\bibinfo{year}{1976}),
  \urlprefix\url{http://link.aps.org/doi/10.1103/PhysRevB.14.1165}.

\bibitem[{\citenamefont{Millis}(1993)}]{PhysRevB.48.7183}
\bibinfo{author}{\bibfnamefont{A.~J.} \bibnamefont{Millis}},
  \bibinfo{journal}{Phys. Rev. B} \textbf{\bibinfo{volume}{48}},
  \bibinfo{pages}{7183} (\bibinfo{year}{1993}),
  \urlprefix\url{http://link.aps.org/doi/10.1103/PhysRevB.48.7183}.

\bibitem[{\citenamefont{L\"ohneysen et~al.}(2007)\citenamefont{L\"ohneysen,
  Rosch, Vojta, and W\"olfle}}]{RevModPhys.79.1015}
\bibinfo{author}{\bibfnamefont{H.~v.} \bibnamefont{L\"ohneysen}},
  \bibinfo{author}{\bibfnamefont{A.}~\bibnamefont{Rosch}},
  \bibinfo{author}{\bibfnamefont{M.}~\bibnamefont{Vojta}}, \bibnamefont{and}
  \bibinfo{author}{\bibfnamefont{P.}~\bibnamefont{W\"olfle}},
  \bibinfo{journal}{Rev. Mod. Phys.} \textbf{\bibinfo{volume}{79}},
  \bibinfo{pages}{1015} (\bibinfo{year}{2007}),
  \urlprefix\url{http://link.aps.org/doi/10.1103/RevModPhys.79.1015}.

\bibitem[{\citenamefont{Stewart}(2001)}]{RevModPhys.73.797}
\bibinfo{author}{\bibfnamefont{G.~R.} \bibnamefont{Stewart}},
  \bibinfo{journal}{Rev. Mod. Phys.} \textbf{\bibinfo{volume}{73}},
  \bibinfo{pages}{797} (\bibinfo{year}{2001}),
  \urlprefix\url{http://link.aps.org/doi/10.1103/RevModPhys.73.797}.

\bibitem[{\citenamefont{Senthil}(2008)}]{PhysRevB.78.035103}
\bibinfo{author}{\bibfnamefont{T.}~\bibnamefont{Senthil}},
  \bibinfo{journal}{Phys. Rev. B} \textbf{\bibinfo{volume}{78}},
  \bibinfo{pages}{035103} (\bibinfo{year}{2008}),
  \urlprefix\url{http://link.aps.org/doi/10.1103/PhysRevB.78.035103}.

\bibitem[{\citenamefont{Holstein et~al.}(1973)\citenamefont{Holstein, Norton,
  and Pincus}}]{PhysRevB.8.2649}
\bibinfo{author}{\bibfnamefont{T.}~\bibnamefont{Holstein}},
  \bibinfo{author}{\bibfnamefont{R.~E.} \bibnamefont{Norton}},
  \bibnamefont{and} \bibinfo{author}{\bibfnamefont{P.}~\bibnamefont{Pincus}},
  \bibinfo{journal}{Phys. Rev. B} \textbf{\bibinfo{volume}{8}},
  \bibinfo{pages}{2649} (\bibinfo{year}{1973}),
  \urlprefix\url{http://link.aps.org/doi/10.1103/PhysRevB.8.2649}.

\bibitem[{\citenamefont{Reizer}(1989)}]{PhysRevB.40.11571}
\bibinfo{author}{\bibfnamefont{M.~Y.} \bibnamefont{Reizer}},
  \bibinfo{journal}{Phys. Rev. B} \textbf{\bibinfo{volume}{40}},
  \bibinfo{pages}{11571} (\bibinfo{year}{1989}),
  \urlprefix\url{http://link.aps.org/doi/10.1103/PhysRevB.40.11571}.

\bibitem[{\citenamefont{Altshuler et~al.}(1994)\citenamefont{Altshuler, Ioffe,
  and Millis}}]{PhysRevB.50.14048}
\bibinfo{author}{\bibfnamefont{B.~L.} \bibnamefont{Altshuler}},
  \bibinfo{author}{\bibfnamefont{L.~B.} \bibnamefont{Ioffe}}, \bibnamefont{and}
  \bibinfo{author}{\bibfnamefont{A.~J.} \bibnamefont{Millis}},
  \bibinfo{journal}{Phys. Rev. B} \textbf{\bibinfo{volume}{50}},
  \bibinfo{pages}{14048} (\bibinfo{year}{1994}),
  \urlprefix\url{http://link.aps.org/doi/10.1103/PhysRevB.50.14048}.

\bibitem[{\citenamefont{Kim et~al.}(1994)\citenamefont{Kim, Furusaki, Wen, and
  Lee}}]{PhysRevB.50.17917}
\bibinfo{author}{\bibfnamefont{Y.~B.} \bibnamefont{Kim}},
  \bibinfo{author}{\bibfnamefont{A.}~\bibnamefont{Furusaki}},
  \bibinfo{author}{\bibfnamefont{X.-G.} \bibnamefont{Wen}}, \bibnamefont{and}
  \bibinfo{author}{\bibfnamefont{P.~A.} \bibnamefont{Lee}},
  \bibinfo{journal}{Phys. Rev. B} \textbf{\bibinfo{volume}{50}},
  \bibinfo{pages}{17917} (\bibinfo{year}{1994}),
  \urlprefix\url{http://link.aps.org/doi/10.1103/PhysRevB.50.17917}.

\bibitem[{\citenamefont{Lee}(1989)}]{PhysRevLett.63.680}
\bibinfo{author}{\bibfnamefont{P.~A.} \bibnamefont{Lee}},
  \bibinfo{journal}{Phys. Rev. Lett.} \textbf{\bibinfo{volume}{63}},
  \bibinfo{pages}{680} (\bibinfo{year}{1989}),
  \urlprefix\url{http://link.aps.org/doi/10.1103/PhysRevLett.63.680}.

\bibitem[{\citenamefont{Polchinski}(1994)}]{polchinski1994low}
\bibinfo{author}{\bibfnamefont{J.}~\bibnamefont{Polchinski}},
  \bibinfo{journal}{Nuclear Physics B} \textbf{\bibinfo{volume}{422}},
  \bibinfo{pages}{617} (\bibinfo{year}{1994}).

\bibitem[{\citenamefont{Lee and Nagaosa}(1992)}]{PhysRevB.46.5621}
\bibinfo{author}{\bibfnamefont{P.~A.} \bibnamefont{Lee}} \bibnamefont{and}
  \bibinfo{author}{\bibfnamefont{N.}~\bibnamefont{Nagaosa}},
  \bibinfo{journal}{Phys. Rev. B} \textbf{\bibinfo{volume}{46}},
  \bibinfo{pages}{5621} (\bibinfo{year}{1992}),
  \urlprefix\url{http://link.aps.org/doi/10.1103/PhysRevB.46.5621}.

\bibitem[{\citenamefont{Nayak and Wilczek}(1994)}]{nayak1994non}
\bibinfo{author}{\bibfnamefont{C.}~\bibnamefont{Nayak}} \bibnamefont{and}
  \bibinfo{author}{\bibfnamefont{F.}~\bibnamefont{Wilczek}},
  \bibinfo{journal}{Nuclear Physics B} \textbf{\bibinfo{volume}{417}},
  \bibinfo{pages}{359} (\bibinfo{year}{1994}).

\bibitem[{\citenamefont{Lee}(2009)}]{PhysRevB.80.165102}
\bibinfo{author}{\bibfnamefont{S.-S.} \bibnamefont{Lee}},
  \bibinfo{journal}{Phys. Rev. B} \textbf{\bibinfo{volume}{80}},
  \bibinfo{pages}{165102} (\bibinfo{year}{2009}),
  \urlprefix\url{http://link.aps.org/doi/10.1103/PhysRevB.80.165102}.

\bibitem[{\citenamefont{Metlitski and
  Sachdev}(2010{\natexlab{a}})}]{PhysRevB.82.075127}
\bibinfo{author}{\bibfnamefont{M.~A.} \bibnamefont{Metlitski}}
  \bibnamefont{and} \bibinfo{author}{\bibfnamefont{S.}~\bibnamefont{Sachdev}},
  \bibinfo{journal}{Phys. Rev. B} \textbf{\bibinfo{volume}{82}},
  \bibinfo{pages}{075127} (\bibinfo{year}{2010}{\natexlab{a}}),
  \urlprefix\url{http://link.aps.org/doi/10.1103/PhysRevB.82.075127}.

\bibitem[{\citenamefont{Mross et~al.}(2010)\citenamefont{Mross, McGreevy, Liu,
  and Senthil}}]{PhysRevB.82.045121}
\bibinfo{author}{\bibfnamefont{D.~F.} \bibnamefont{Mross}},
  \bibinfo{author}{\bibfnamefont{J.}~\bibnamefont{McGreevy}},
  \bibinfo{author}{\bibfnamefont{H.}~\bibnamefont{Liu}}, \bibnamefont{and}
  \bibinfo{author}{\bibfnamefont{T.}~\bibnamefont{Senthil}},
  \bibinfo{journal}{Phys. Rev. B} \textbf{\bibinfo{volume}{82}},
  \bibinfo{pages}{045121} (\bibinfo{year}{2010}),
  \urlprefix\url{http://link.aps.org/doi/10.1103/PhysRevB.82.045121}.

\bibitem[{\citenamefont{Jiang et~al.}(2013)\citenamefont{Jiang, Block,
  Mishmash, Garrison, Sheng, Motrunich, and Fisher}}]{jiang2013non}
\bibinfo{author}{\bibfnamefont{H.-C.} \bibnamefont{Jiang}},
  \bibinfo{author}{\bibfnamefont{M.~S.} \bibnamefont{Block}},
  \bibinfo{author}{\bibfnamefont{R.~V.} \bibnamefont{Mishmash}},
  \bibinfo{author}{\bibfnamefont{J.~R.} \bibnamefont{Garrison}},
  \bibinfo{author}{\bibfnamefont{D.}~\bibnamefont{Sheng}},
  \bibinfo{author}{\bibfnamefont{O.~I.} \bibnamefont{Motrunich}},
  \bibnamefont{and} \bibinfo{author}{\bibfnamefont{M.~P.}
  \bibnamefont{Fisher}}, \bibinfo{journal}{Nature}
  \textbf{\bibinfo{volume}{493}}, \bibinfo{pages}{39} (\bibinfo{year}{2013}).

\bibitem[{\citenamefont{Dalidovich and Lee}(2013)}]{PhysRevB.88.245106}
\bibinfo{author}{\bibfnamefont{D.}~\bibnamefont{Dalidovich}} \bibnamefont{and}
  \bibinfo{author}{\bibfnamefont{S.-S.} \bibnamefont{Lee}},
  \bibinfo{journal}{Phys. Rev. B} \textbf{\bibinfo{volume}{88}},
  \bibinfo{pages}{245106} (\bibinfo{year}{2013}),
  \urlprefix\url{http://link.aps.org/doi/10.1103/PhysRevB.88.245106}.

\bibitem[{\citenamefont{Sur and Lee}(2014)}]{PhysRevB.90.045121}
\bibinfo{author}{\bibfnamefont{S.}~\bibnamefont{Sur}} \bibnamefont{and}
  \bibinfo{author}{\bibfnamefont{S.-S.} \bibnamefont{Lee}},
  \bibinfo{journal}{Phys. Rev. B} \textbf{\bibinfo{volume}{90}},
  \bibinfo{pages}{045121} (\bibinfo{year}{2014}),
  \urlprefix\url{http://link.aps.org/doi/10.1103/PhysRevB.90.045121}.

\bibitem[{\citenamefont{Holder and Metzner}(2015)}]{PhysRevB.92.041112}
\bibinfo{author}{\bibfnamefont{T.}~\bibnamefont{Holder}} \bibnamefont{and}
  \bibinfo{author}{\bibfnamefont{W.}~\bibnamefont{Metzner}},
  \bibinfo{journal}{Phys. Rev. B} \textbf{\bibinfo{volume}{92}},
  \bibinfo{pages}{041112} (\bibinfo{year}{2015}),
  \urlprefix\url{http://link.aps.org/doi/10.1103/PhysRevB.92.041112}.

\bibitem[{\citenamefont{Helm et~al.}(2010)\citenamefont{Helm, Kartsovnik,
  Sheikin, Bartkowiak, Wolff-Fabris, Bittner, Biberacher, Lambacher, Erb,
  Wosnitza et~al.}}]{PhysRevLett.105.247002}
\bibinfo{author}{\bibfnamefont{T.}~\bibnamefont{Helm}},
  \bibinfo{author}{\bibfnamefont{M.~V.} \bibnamefont{Kartsovnik}},
  \bibinfo{author}{\bibfnamefont{I.}~\bibnamefont{Sheikin}},
  \bibinfo{author}{\bibfnamefont{M.}~\bibnamefont{Bartkowiak}},
  \bibinfo{author}{\bibfnamefont{F.}~\bibnamefont{Wolff-Fabris}},
  \bibinfo{author}{\bibfnamefont{N.}~\bibnamefont{Bittner}},
  \bibinfo{author}{\bibfnamefont{W.}~\bibnamefont{Biberacher}},
  \bibinfo{author}{\bibfnamefont{M.}~\bibnamefont{Lambacher}},
  \bibinfo{author}{\bibfnamefont{A.}~\bibnamefont{Erb}},
  \bibinfo{author}{\bibfnamefont{J.}~\bibnamefont{Wosnitza}},
  \bibnamefont{et~al.}, \bibinfo{journal}{Phys. Rev. Lett.}
  \textbf{\bibinfo{volume}{105}}, \bibinfo{pages}{247002}
  (\bibinfo{year}{2010}),
  \urlprefix\url{http://link.aps.org/doi/10.1103/PhysRevLett.105.247002}.

\bibitem[{\citenamefont{Hashimoto et~al.}(2012)\citenamefont{Hashimoto, Cho,
  Shibauchi, Kasahara, Mizukami, Katsumata, Tsuruhara, Terashima, Ikeda,
  Tanatar et~al.}}]{Hashimoto22062012}
\bibinfo{author}{\bibfnamefont{K.}~\bibnamefont{Hashimoto}},
  \bibinfo{author}{\bibfnamefont{K.}~\bibnamefont{Cho}},
  \bibinfo{author}{\bibfnamefont{T.}~\bibnamefont{Shibauchi}},
  \bibinfo{author}{\bibfnamefont{S.}~\bibnamefont{Kasahara}},
  \bibinfo{author}{\bibfnamefont{Y.}~\bibnamefont{Mizukami}},
  \bibinfo{author}{\bibfnamefont{R.}~\bibnamefont{Katsumata}},
  \bibinfo{author}{\bibfnamefont{Y.}~\bibnamefont{Tsuruhara}},
  \bibinfo{author}{\bibfnamefont{T.}~\bibnamefont{Terashima}},
  \bibinfo{author}{\bibfnamefont{H.}~\bibnamefont{Ikeda}},
  \bibinfo{author}{\bibfnamefont{M.~A.} \bibnamefont{Tanatar}},
  \bibnamefont{et~al.}, \bibinfo{journal}{Science}
  \textbf{\bibinfo{volume}{336}}, \bibinfo{pages}{1554} (\bibinfo{year}{2012}),
  \eprint{http://www.sciencemag.org/content/336/6088/1554.full.pdf},
  \urlprefix\url{http://www.sciencemag.org/content/336/6088/1554.abstract}.

\bibitem[{\citenamefont{Park et~al.}(2006)\citenamefont{Park, Ronning, Yuan,
  Salamon, Movshovich, Sarrao, and Thompson}}]{park2006hidden}
\bibinfo{author}{\bibfnamefont{T.}~\bibnamefont{Park}},
  \bibinfo{author}{\bibfnamefont{F.}~\bibnamefont{Ronning}},
  \bibinfo{author}{\bibfnamefont{H.}~\bibnamefont{Yuan}},
  \bibinfo{author}{\bibfnamefont{M.}~\bibnamefont{Salamon}},
  \bibinfo{author}{\bibfnamefont{R.}~\bibnamefont{Movshovich}},
  \bibinfo{author}{\bibfnamefont{J.}~\bibnamefont{Sarrao}}, \bibnamefont{and}
  \bibinfo{author}{\bibfnamefont{J.}~\bibnamefont{Thompson}},
  \bibinfo{journal}{Nature} \textbf{\bibinfo{volume}{440}}, \bibinfo{pages}{65}
  (\bibinfo{year}{2006}).

\bibitem[{\citenamefont{Abanov and Chubukov}(2000)}]{PhysRevLett.84.5608}
\bibinfo{author}{\bibfnamefont{A.}~\bibnamefont{Abanov}} \bibnamefont{and}
  \bibinfo{author}{\bibfnamefont{A.~V.} \bibnamefont{Chubukov}},
  \bibinfo{journal}{Phys. Rev. Lett.} \textbf{\bibinfo{volume}{84}},
  \bibinfo{pages}{5608} (\bibinfo{year}{2000}),
  \urlprefix\url{http://link.aps.org/doi/10.1103/PhysRevLett.84.5608}.

\bibitem[{\citenamefont{Abanov et~al.}(2003)\citenamefont{Abanov, Chubukov, and
  Schmalian}}]{doi:10.1080/0001873021000057123}
\bibinfo{author}{\bibfnamefont{A.}~\bibnamefont{Abanov}},
  \bibinfo{author}{\bibfnamefont{A.~V.} \bibnamefont{Chubukov}},
  \bibnamefont{and}
  \bibinfo{author}{\bibfnamefont{J.}~\bibnamefont{Schmalian}},
  \bibinfo{journal}{Advances in Physics} \textbf{\bibinfo{volume}{52}},
  \bibinfo{pages}{119} (\bibinfo{year}{2003}),
  \eprint{http://www.tandfonline.com/doi/pdf/10.1080/0001873021000057123},
  \urlprefix\url{http://www.tandfonline.com/doi/abs/10.1080/0001873021000057123}.

\bibitem[{\citenamefont{Abanov and Chubukov}(2004)}]{PhysRevLett.93.255702}
\bibinfo{author}{\bibfnamefont{A.}~\bibnamefont{Abanov}} \bibnamefont{and}
  \bibinfo{author}{\bibfnamefont{A.}~\bibnamefont{Chubukov}},
  \bibinfo{journal}{Phys. Rev. Lett.} \textbf{\bibinfo{volume}{93}},
  \bibinfo{pages}{255702} (\bibinfo{year}{2004}),
  \urlprefix\url{http://link.aps.org/doi/10.1103/PhysRevLett.93.255702}.

\bibitem[{\citenamefont{Metlitski and
  Sachdev}(2010{\natexlab{b}})}]{PhysRevB.82.075128}
\bibinfo{author}{\bibfnamefont{M.~A.} \bibnamefont{Metlitski}}
  \bibnamefont{and} \bibinfo{author}{\bibfnamefont{S.}~\bibnamefont{Sachdev}},
  \bibinfo{journal}{Phys. Rev. B} \textbf{\bibinfo{volume}{82}},
  \bibinfo{pages}{075128} (\bibinfo{year}{2010}{\natexlab{b}}),
  \urlprefix\url{http://link.aps.org/doi/10.1103/PhysRevB.82.075128}.

\bibitem[{\citenamefont{Abrahams and Wolfle}(2012)}]{Abrahams28022012}
\bibinfo{author}{\bibfnamefont{E.}~\bibnamefont{Abrahams}} \bibnamefont{and}
  \bibinfo{author}{\bibfnamefont{P.}~\bibnamefont{Wolfle}},
  \bibinfo{journal}{Proceedings of the National Academy of Sciences}
  \textbf{\bibinfo{volume}{109}}, \bibinfo{pages}{3238} (\bibinfo{year}{2012}),
  \eprint{http://www.pnas.org/content/109/9/3238.full.pdf},
  \urlprefix\url{http://www.pnas.org/content/109/9/3238.abstract}.

\bibitem[{\citenamefont{Sur and Lee}(2015)}]{PhysRevB.91.125136}
\bibinfo{author}{\bibfnamefont{S.}~\bibnamefont{Sur}} \bibnamefont{and}
  \bibinfo{author}{\bibfnamefont{S.-S.} \bibnamefont{Lee}},
  \bibinfo{journal}{Phys. Rev. B} \textbf{\bibinfo{volume}{91}},
  \bibinfo{pages}{125136} (\bibinfo{year}{2015}),
  \urlprefix\url{http://link.aps.org/doi/10.1103/PhysRevB.91.125136}.

\bibitem[{\citenamefont{Maier and Strack}(2016)}]{PhysRevB.93.165114}
\bibinfo{author}{\bibfnamefont{S.~A.} \bibnamefont{Maier}} \bibnamefont{and}
  \bibinfo{author}{\bibfnamefont{P.}~\bibnamefont{Strack}},
  \bibinfo{journal}{Phys. Rev. B} \textbf{\bibinfo{volume}{93}},
  \bibinfo{pages}{165114} (\bibinfo{year}{2016}),
  \urlprefix\url{http://link.aps.org/doi/10.1103/PhysRevB.93.165114}.

\bibitem[{\citenamefont{Berg et~al.}(2012)\citenamefont{Berg, Metlitski, and
  Sachdev}}]{Berg21122012}
\bibinfo{author}{\bibfnamefont{E.}~\bibnamefont{Berg}},
  \bibinfo{author}{\bibfnamefont{M.~A.} \bibnamefont{Metlitski}},
  \bibnamefont{and} \bibinfo{author}{\bibfnamefont{S.}~\bibnamefont{Sachdev}},
  \bibinfo{journal}{Science} \textbf{\bibinfo{volume}{338}},
  \bibinfo{pages}{1606} (\bibinfo{year}{2012}),
  \eprint{http://www.sciencemag.org/content/338/6114/1606.full.pdf},
  \urlprefix\url{http://www.sciencemag.org/content/338/6114/1606.abstract}.

\bibitem[{\citenamefont{{Schattner} et~al.}(2015)\citenamefont{{Schattner},
  {Gerlach}, {Trebst}, and {Berg}}}]{2015arXiv151207257S}
\bibinfo{author}{\bibfnamefont{Y.}~\bibnamefont{{Schattner}}},
  \bibinfo{author}{\bibfnamefont{M.~H.} \bibnamefont{{Gerlach}}},
  \bibinfo{author}{\bibfnamefont{S.}~\bibnamefont{{Trebst}}}, \bibnamefont{and}
  \bibinfo{author}{\bibfnamefont{E.}~\bibnamefont{{Berg}}},
  \bibinfo{journal}{ArXiv e-prints}  (\bibinfo{year}{2015}),
  \eprint{1512.07257}.

\bibitem[{\citenamefont{Fitzpatrick et~al.}(2013)\citenamefont{Fitzpatrick,
  Kachru, Kaplan, and Raghu}}]{PhysRevB.88.125116}
\bibinfo{author}{\bibfnamefont{A.~L.} \bibnamefont{Fitzpatrick}},
  \bibinfo{author}{\bibfnamefont{S.}~\bibnamefont{Kachru}},
  \bibinfo{author}{\bibfnamefont{J.}~\bibnamefont{Kaplan}}, \bibnamefont{and}
  \bibinfo{author}{\bibfnamefont{S.}~\bibnamefont{Raghu}},
  \bibinfo{journal}{Phys. Rev. B} \textbf{\bibinfo{volume}{88}},
  \bibinfo{pages}{125116} (\bibinfo{year}{2013}),
  \urlprefix\url{http://link.aps.org/doi/10.1103/PhysRevB.88.125116}.

\bibitem[{\citenamefont{{Lunts} et~al.}(2017)\citenamefont{{Lunts}, {Schlief},
  and {Lee}}}]{2017arXiv170108218L}
\bibinfo{author}{\bibfnamefont{P.}~\bibnamefont{{Lunts}}},
  \bibinfo{author}{\bibfnamefont{A.}~\bibnamefont{{Schlief}}},
  \bibnamefont{and} \bibinfo{author}{\bibfnamefont{S.-S.} \bibnamefont{{Lee}}},
  \bibinfo{journal}{ArXiv e-prints}  (\bibinfo{year}{2017}),
  \eprint{1701.08218}.

\bibitem[{\citenamefont{Huh and Sachdev}(2008)}]{PhysRevB.78.064512}
\bibinfo{author}{\bibfnamefont{Y.}~\bibnamefont{Huh}} \bibnamefont{and}
  \bibinfo{author}{\bibfnamefont{S.}~\bibnamefont{Sachdev}},
  \bibinfo{journal}{Phys. Rev. B} \textbf{\bibinfo{volume}{78}},
  \bibinfo{pages}{064512} (\bibinfo{year}{2008}),
  \urlprefix\url{http://link.aps.org/doi/10.1103/PhysRevB.78.064512}.

\bibitem[{\citenamefont{Varma}(2015)}]{PhysRevLett.115.186405}
\bibinfo{author}{\bibfnamefont{C.~M.} \bibnamefont{Varma}},
  \bibinfo{journal}{Phys. Rev. Lett.} \textbf{\bibinfo{volume}{115}},
  \bibinfo{pages}{186405} (\bibinfo{year}{2015}),
  \urlprefix\url{http://link.aps.org/doi/10.1103/PhysRevLett.115.186405}.

\bibitem[{\citenamefont{Varma et~al.}(1989)\citenamefont{Varma, Littlewood,
  Schmitt-Rink, Abrahams, and Ruckenstein}}]{varma1989phenomenology}
\bibinfo{author}{\bibfnamefont{C.}~\bibnamefont{Varma}},
  \bibinfo{author}{\bibfnamefont{P.~B.} \bibnamefont{Littlewood}},
  \bibinfo{author}{\bibfnamefont{S.}~\bibnamefont{Schmitt-Rink}},
  \bibinfo{author}{\bibfnamefont{E.}~\bibnamefont{Abrahams}}, \bibnamefont{and}
  \bibinfo{author}{\bibfnamefont{A.}~\bibnamefont{Ruckenstein}},
  \bibinfo{journal}{Physical Review Letters} \textbf{\bibinfo{volume}{63}},
  \bibinfo{pages}{1996} (\bibinfo{year}{1989}).

\bibitem[{\citenamefont{Luttinger and Ward}(1960)}]{PhysRev.118.1417}
\bibinfo{author}{\bibfnamefont{J.~M.} \bibnamefont{Luttinger}}
  \bibnamefont{and} \bibinfo{author}{\bibfnamefont{J.~C.} \bibnamefont{Ward}},
  \bibinfo{journal}{Phys. Rev.} \textbf{\bibinfo{volume}{118}},
  \bibinfo{pages}{1417} (\bibinfo{year}{1960}),
  \urlprefix\url{http://link.aps.org/doi/10.1103/PhysRev.118.1417}.

\bibitem[{\citenamefont{Patel et~al.}(2015)\citenamefont{Patel, Strack, and
  Sachdev}}]{PhysRevB.92.165105}
\bibinfo{author}{\bibfnamefont{A.~A.} \bibnamefont{Patel}},
  \bibinfo{author}{\bibfnamefont{P.}~\bibnamefont{Strack}}, \bibnamefont{and}
  \bibinfo{author}{\bibfnamefont{S.}~\bibnamefont{Sachdev}},
  \bibinfo{journal}{Phys. Rev. B} \textbf{\bibinfo{volume}{92}},
  \bibinfo{pages}{165105} (\bibinfo{year}{2015}),
  \urlprefix\url{http://link.aps.org/doi/10.1103/PhysRevB.92.165105}.

\bibitem[{\citenamefont{Scalapino et~al.}(1986)\citenamefont{Scalapino, Loh~Jr,
  and Hirsch}}]{scalapino1986d}
\bibinfo{author}{\bibfnamefont{D.}~\bibnamefont{Scalapino}},
  \bibinfo{author}{\bibfnamefont{E.}~\bibnamefont{Loh~Jr}}, \bibnamefont{and}
  \bibinfo{author}{\bibfnamefont{J.}~\bibnamefont{Hirsch}},
  \bibinfo{journal}{Physical Review B} \textbf{\bibinfo{volume}{34}},
  \bibinfo{pages}{8190} (\bibinfo{year}{1986}).

\bibitem[{\citenamefont{Miyake et~al.}(1986)\citenamefont{Miyake, Schmitt-Rink,
  and Varma}}]{PhysRevB.34.6554}
\bibinfo{author}{\bibfnamefont{K.}~\bibnamefont{Miyake}},
  \bibinfo{author}{\bibfnamefont{S.}~\bibnamefont{Schmitt-Rink}},
  \bibnamefont{and} \bibinfo{author}{\bibfnamefont{C.~M.} \bibnamefont{Varma}},
  \bibinfo{journal}{Phys. Rev. B} \textbf{\bibinfo{volume}{34}},
  \bibinfo{pages}{6554} (\bibinfo{year}{1986}),
  \urlprefix\url{http://link.aps.org/doi/10.1103/PhysRevB.34.6554}.

\bibitem[{\citenamefont{Moriya et~al.}(1990)\citenamefont{Moriya, Takahashi,
  and Ueda}}]{doi:10.1143/JPSJ.59.2905}
\bibinfo{author}{\bibfnamefont{T.}~\bibnamefont{Moriya}},
  \bibinfo{author}{\bibfnamefont{Y.}~\bibnamefont{Takahashi}},
  \bibnamefont{and} \bibinfo{author}{\bibfnamefont{K.}~\bibnamefont{Ueda}},
  \bibinfo{journal}{Journal of the Physical Society of Japan}
  \textbf{\bibinfo{volume}{59}}, \bibinfo{pages}{2905} (\bibinfo{year}{1990}),
  \eprint{http://dx.doi.org/10.1143/JPSJ.59.2905},
  \urlprefix\url{http://dx.doi.org/10.1143/JPSJ.59.2905}.

\bibitem[{\citenamefont{{Li} et~al.}(2015)\citenamefont{{Li}, {Wang}, {Yao},
  and {Lee}}}]{2015arXiv151204541L}
\bibinfo{author}{\bibfnamefont{Z.-X.} \bibnamefont{{Li}}},
  \bibinfo{author}{\bibfnamefont{F.}~\bibnamefont{{Wang}}},
  \bibinfo{author}{\bibfnamefont{H.}~\bibnamefont{{Yao}}}, \bibnamefont{and}
  \bibinfo{author}{\bibfnamefont{D.-H.} \bibnamefont{{Lee}}},
  \bibinfo{journal}{ArXiv e-prints}  (\bibinfo{year}{2015}),
  \eprint{1512.04541}.

\bibitem[{\citenamefont{Son}(1999)}]{PhysRevD.59.094019}
\bibinfo{author}{\bibfnamefont{D.~T.} \bibnamefont{Son}},
  \bibinfo{journal}{Phys. Rev. D} \textbf{\bibinfo{volume}{59}},
  \bibinfo{pages}{094019} (\bibinfo{year}{1999}),
  \urlprefix\url{http://link.aps.org/doi/10.1103/PhysRevD.59.094019}.

\bibitem[{\citenamefont{Monthoux et~al.}(1992)\citenamefont{Monthoux, Balatsky,
  and Pines}}]{PhysRevB.46.14803}
\bibinfo{author}{\bibfnamefont{P.}~\bibnamefont{Monthoux}},
  \bibinfo{author}{\bibfnamefont{A.~V.} \bibnamefont{Balatsky}},
  \bibnamefont{and} \bibinfo{author}{\bibfnamefont{D.}~\bibnamefont{Pines}},
  \bibinfo{journal}{Phys. Rev. B} \textbf{\bibinfo{volume}{46}},
  \bibinfo{pages}{14803} (\bibinfo{year}{1992}),
  \urlprefix\url{http://link.aps.org/doi/10.1103/PhysRevB.46.14803}.

\bibitem[{\citenamefont{Chubukov and Schmalian}(2005)}]{PhysRevB.72.174520}
\bibinfo{author}{\bibfnamefont{A.~V.} \bibnamefont{Chubukov}} \bibnamefont{and}
  \bibinfo{author}{\bibfnamefont{J.}~\bibnamefont{Schmalian}},
  \bibinfo{journal}{Phys. Rev. B} \textbf{\bibinfo{volume}{72}},
  \bibinfo{pages}{174520} (\bibinfo{year}{2005}),
  \urlprefix\url{http://link.aps.org/doi/10.1103/PhysRevB.72.174520}.

\bibitem[{\citenamefont{Metlitski et~al.}(2015)\citenamefont{Metlitski, Mross,
  Sachdev, and Senthil}}]{PhysRevB.91.115111}
\bibinfo{author}{\bibfnamefont{M.~A.} \bibnamefont{Metlitski}},
  \bibinfo{author}{\bibfnamefont{D.~F.} \bibnamefont{Mross}},
  \bibinfo{author}{\bibfnamefont{S.}~\bibnamefont{Sachdev}}, \bibnamefont{and}
  \bibinfo{author}{\bibfnamefont{T.}~\bibnamefont{Senthil}},
  \bibinfo{journal}{Phys. Rev. B} \textbf{\bibinfo{volume}{91}},
  \bibinfo{pages}{115111} (\bibinfo{year}{2015}),
  \urlprefix\url{http://link.aps.org/doi/10.1103/PhysRevB.91.115111}.

\bibitem[{\citenamefont{Lederer et~al.}(2015)\citenamefont{Lederer, Schattner,
  Berg, and Kivelson}}]{PhysRevLett.114.097001}
\bibinfo{author}{\bibfnamefont{S.}~\bibnamefont{Lederer}},
  \bibinfo{author}{\bibfnamefont{Y.}~\bibnamefont{Schattner}},
  \bibinfo{author}{\bibfnamefont{E.}~\bibnamefont{Berg}}, \bibnamefont{and}
  \bibinfo{author}{\bibfnamefont{S.~A.} \bibnamefont{Kivelson}},
  \bibinfo{journal}{Phys. Rev. Lett.} \textbf{\bibinfo{volume}{114}},
  \bibinfo{pages}{097001} (\bibinfo{year}{2015}),
  \urlprefix\url{http://link.aps.org/doi/10.1103/PhysRevLett.114.097001}.

\bibitem[{\citenamefont{Horio et~al.}(2016)\citenamefont{Horio, Adachi, Mori,
  Takahashi, Yoshida, Suzuki, Ambolode~II, Okazaki, Ono, Kumigashira
  et~al.}}]{Horio2016}
\bibinfo{author}{\bibfnamefont{M.}~\bibnamefont{Horio}},
  \bibinfo{author}{\bibfnamefont{T.}~\bibnamefont{Adachi}},
  \bibinfo{author}{\bibfnamefont{Y.}~\bibnamefont{Mori}},
  \bibinfo{author}{\bibfnamefont{A.}~\bibnamefont{Takahashi}},
  \bibinfo{author}{\bibfnamefont{T.}~\bibnamefont{Yoshida}},
  \bibinfo{author}{\bibfnamefont{H.}~\bibnamefont{Suzuki}},
  \bibinfo{author}{\bibfnamefont{L.~C.~C.} \bibnamefont{Ambolode~II}},
  \bibinfo{author}{\bibfnamefont{K.}~\bibnamefont{Okazaki}},
  \bibinfo{author}{\bibfnamefont{K.}~\bibnamefont{Ono}},
  \bibinfo{author}{\bibfnamefont{H.}~\bibnamefont{Kumigashira}},
  \bibnamefont{et~al.}, \bibinfo{journal}{Nature Communications}
  \textbf{\bibinfo{volume}{7}}, \bibinfo{pages}{10567 EP }
  (\bibinfo{year}{2016}), \bibinfo{note}{article},
  \urlprefix\url{http://dx.doi.org/10.1038/ncomms10567}.

\bibitem[{\citenamefont{Sur and Lee}(2016)}]{PhysRevB.94.195135}
\bibinfo{author}{\bibfnamefont{S.}~\bibnamefont{Sur}} \bibnamefont{and}
  \bibinfo{author}{\bibfnamefont{S.-S.} \bibnamefont{Lee}},
  \bibinfo{journal}{Phys. Rev. B} \textbf{\bibinfo{volume}{94}},
  \bibinfo{pages}{195135} (\bibinfo{year}{2016}),
  \urlprefix\url{http://link.aps.org/doi/10.1103/PhysRevB.94.195135}.

\bibitem[{\citenamefont{Wilson et~al.}(2006)\citenamefont{Wilson, Dai, Li, Chi,
  Kang, and Lynn}}]{Wilson2006}
\bibinfo{author}{\bibfnamefont{S.~D.} \bibnamefont{Wilson}},
  \bibinfo{author}{\bibfnamefont{P.}~\bibnamefont{Dai}},
  \bibinfo{author}{\bibfnamefont{S.}~\bibnamefont{Li}},
  \bibinfo{author}{\bibfnamefont{S.}~\bibnamefont{Chi}},
  \bibinfo{author}{\bibfnamefont{H.~J.} \bibnamefont{Kang}}, \bibnamefont{and}
  \bibinfo{author}{\bibfnamefont{J.~W.} \bibnamefont{Lynn}},
  \bibinfo{journal}{Nature} \textbf{\bibinfo{volume}{442}}, \bibinfo{pages}{59}
  (\bibinfo{year}{2006}), ISSN \bibinfo{issn}{0028-0836},
  \urlprefix\url{http://dx.doi.org/10.1038/nature04857}.

\bibitem[{\citenamefont{Motoyama et~al.}(2007)\citenamefont{Motoyama, Yu,
  Vishik, Vajk, Mang, and Greven}}]{Motoyama2007}
\bibinfo{author}{\bibfnamefont{E.~M.} \bibnamefont{Motoyama}},
  \bibinfo{author}{\bibfnamefont{G.}~\bibnamefont{Yu}},
  \bibinfo{author}{\bibfnamefont{I.~M.} \bibnamefont{Vishik}},
  \bibinfo{author}{\bibfnamefont{O.~P.} \bibnamefont{Vajk}},
  \bibinfo{author}{\bibfnamefont{P.~K.} \bibnamefont{Mang}}, \bibnamefont{and}
  \bibinfo{author}{\bibfnamefont{M.}~\bibnamefont{Greven}},
  \bibinfo{journal}{Nature} \textbf{\bibinfo{volume}{445}},
  \bibinfo{pages}{186} (\bibinfo{year}{2007}), ISSN \bibinfo{issn}{0028-0836},
  \urlprefix\url{http://dx.doi.org/10.1038/nature05437}.

\bibitem[{\citenamefont{Armitage et~al.}(2001)\citenamefont{Armitage, Lu, Kim,
  Damascelli, Shen, Ronning, Feng, Bogdanov, Shen, Onose
  et~al.}}]{PhysRevLett.87.147003}
\bibinfo{author}{\bibfnamefont{N.~P.} \bibnamefont{Armitage}},
  \bibinfo{author}{\bibfnamefont{D.~H.} \bibnamefont{Lu}},
  \bibinfo{author}{\bibfnamefont{C.}~\bibnamefont{Kim}},
  \bibinfo{author}{\bibfnamefont{A.}~\bibnamefont{Damascelli}},
  \bibinfo{author}{\bibfnamefont{K.~M.} \bibnamefont{Shen}},
  \bibinfo{author}{\bibfnamefont{F.}~\bibnamefont{Ronning}},
  \bibinfo{author}{\bibfnamefont{D.~L.} \bibnamefont{Feng}},
  \bibinfo{author}{\bibfnamefont{P.}~\bibnamefont{Bogdanov}},
  \bibinfo{author}{\bibfnamefont{Z.-X.} \bibnamefont{Shen}},
  \bibinfo{author}{\bibfnamefont{Y.}~\bibnamefont{Onose}},
  \bibnamefont{et~al.}, \bibinfo{journal}{Phys. Rev. Lett.}
  \textbf{\bibinfo{volume}{87}}, \bibinfo{pages}{147003}
  (\bibinfo{year}{2001}),
  \urlprefix\url{http://link.aps.org/doi/10.1103/PhysRevLett.87.147003}.

\bibitem[{\citenamefont{Schmitt et~al.}(2008)\citenamefont{Schmitt, Lee, Lu,
  Meevasana, Motoyama, Greven, and Shen}}]{PhysRevB.78.100505}
\bibinfo{author}{\bibfnamefont{F.}~\bibnamefont{Schmitt}},
  \bibinfo{author}{\bibfnamefont{W.~S.} \bibnamefont{Lee}},
  \bibinfo{author}{\bibfnamefont{D.-H.} \bibnamefont{Lu}},
  \bibinfo{author}{\bibfnamefont{W.}~\bibnamefont{Meevasana}},
  \bibinfo{author}{\bibfnamefont{E.}~\bibnamefont{Motoyama}},
  \bibinfo{author}{\bibfnamefont{M.}~\bibnamefont{Greven}}, \bibnamefont{and}
  \bibinfo{author}{\bibfnamefont{Z.-X.} \bibnamefont{Shen}},
  \bibinfo{journal}{Phys. Rev. B} \textbf{\bibinfo{volume}{78}},
  \bibinfo{pages}{100505} (\bibinfo{year}{2008}),
  \urlprefix\url{http://link.aps.org/doi/10.1103/PhysRevB.78.100505}.

\end{thebibliography}

\section{Acknowledgment}

We thank 
Andrey Chubukov, 
Patrick Lee, 
Max Metlitski, 
Subir Sachdev
and T. Senthil for comments.
The research was supported in part by 
the Natural Sciences and Engineering Research Council of Canada.
Research at the Perimeter Institute is supported 
in part by the Government of Canada 
through Industry Canada, 
and by the Province of Ontario through the
Ministry of Research and Information.

A. S. and P. L. contributed equally to this work.
Correspondence should be addressed to slee@mcmaster.ca.


\newpage
\appendix

\setcounter{page}{1}
\setcounter{figure}{0}
\setcounter{subfigure}{0}
\setcounter{table}{0}
\setcounter{equation}{0}

\renewcommand\thefigure{A\arabic{figure}}
\renewcommand\thetable{A\arabic{table}}
\renewcommand\theequation{A\arabic{equation}}






\section{Proof of the upper bound for general diagrams}
	\label{sec:HL}

In this section, we prove the upper bound in \eq{EST},
assuming that the fully dressed boson propagator
is given by Eqs. (\ref{eq:D}) and (\ref{eq:cv}) in the small $v$ limit. 
Since the boson propagator is already fully dressed,
we do not need to consider boson self-energy corrections within diagrams. 
The magnitude of a diagram is not simply determined by the number of vertices
because in the small $v$ limit
patches of the Fermi surface become locally nested, 
and the collective mode loses its dispersion. 
When a loop is formed out of dispersionless bosons and nested fermions,
the loop momentum along the Fermi surface becomes unbounded.
For small but nonzero $v$ and $c$,
the divergent integral is cut off 
by a scale which is proportional to $1/v$ or $1/c$.
This gives rise to enhancement factors of $1/v$ or $1/c$.
Our goal is to compute the upper bound of the enhancement factors for general diagrams. 
A diagram is maximally enhanced 
when all the patches of the Fermi surface 
involved in the diagram are nested.
Since the patches are nested pairwise ($1,3$ and $2,4$) in the small $v$ limit,
it is enough to consider diagrams that are made of patches $1,3$
to compute the upper bound without loss of generality.
Diagrams which involve all four patches are generally smaller in magnitude 
than those that involve only $1,3$ or $2,4$ for fixed $L, L_f, E$,
where $L$ is the total number of loops,
$L_f$ is the number of fermion loops
and $E$ is the number of external legs.
We first show that \eq{EST} holds 
for an example 
to illustrate the idea
that is used for a general proof
in the following subsection.

\subsection{Example}

 \begin{figure}[h]
 	\centering
 	\subfigure[]
 	{\includegraphics[scale=0.80]{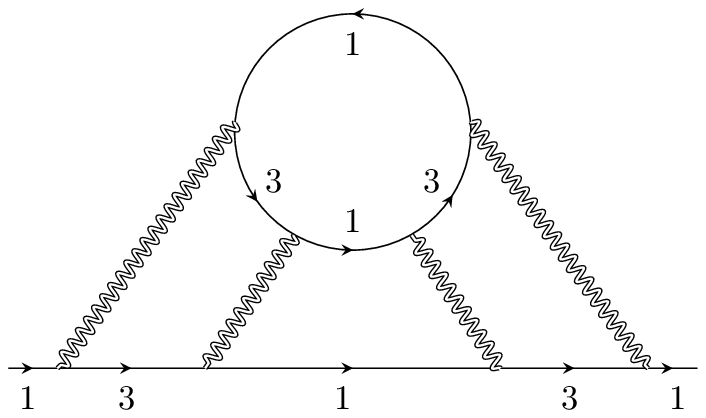}
 		\label{fig:HL1}} ~~~~~
 	\subfigure[]
 	{\includegraphics[scale=0.80]{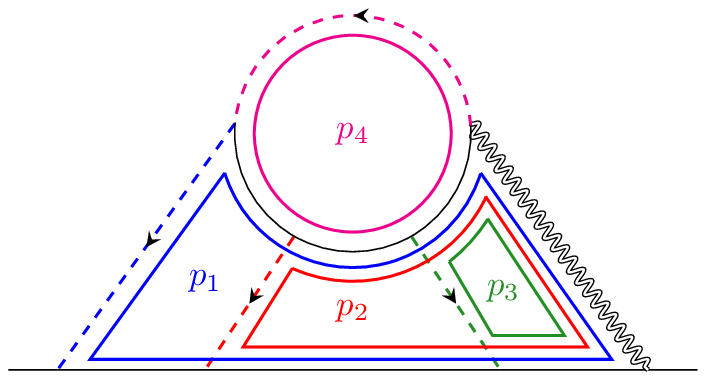}
 		\label{fig:HL2}} ~~~~~
 	\subfigure[]
 	{\includegraphics[scale=0.80]{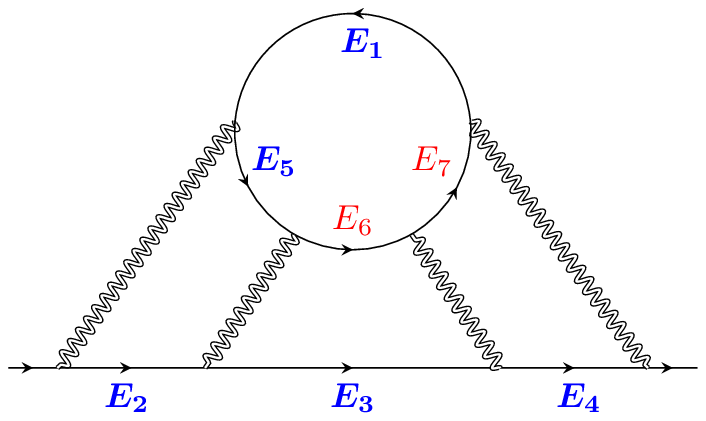}
 		\label{fig:HL3}} ~~~~~
 	\caption{
 		(a) A four-loop diagram with one fermion loop.
 		The numbers next to the fermion lines represent the patch indices.
 		(b) The four exclusive propagators are denoted as dashed lines. The remaining propagators represent the connected tree diagram. Loops (thick solid colored lines) are chosen such that each loop momentum goes through only one of the exclusive propagators.
 		(c) The seven internal fermion propagators whose energies
 		are denoted as $E_l$ with $1 \leq l \leq 7$.
 		$E_1, E_2, .., E_5$ are used as 
 		new integration variables 
 		along with $p_{i}^{'} = c p_{i,x}$ with $i=1,2,3$,
 		as discussed in the text.
 	}
 	\label{fig:HLE}
 \end{figure}

The diagram in \fig{fig:HL1} is a fermion self-energy with one fermion loop and three other loops, 
which we call `mixed loops'. 
For simplicity, we set the external momentum to zero. 
This does not affect the enhancement factors of $1/c$ and $1/v$ 
which originate from large internal momenta.
We label the loop momenta as shown in \fig{fig:HL2}. 
With this choice, 
each mixed loop momentum $p_i$ with $i = 1,2,3$ has a boson line that carries only $p_i$, 
and the fermion loop momentum $p_4$ has a fermion line that carries only $p_4$. 
These four propagators, denoted in \fig{fig:HL2} by dashed lines, are called `exclusive propagators'. 
In the next section we show that it is always possible 
to find such exclusive propagators for every loop momentum in a general diagram. 
The diagram in \fig{fig:HL1} is written as
\begin{eqnarray}
\nonumber
I &\sim& v^{4}
\int 
\prod_{r=1}^4 d p_r \, 
\left( \prod_{j=1}^3 \frac{1}{ |p_{j,0}| + c ( |p_{j,x}| + |p_{j,y}| ) }  \right)
\times \\ && \nonumber 
\frac{1}{ |p_{1,0} + p_{2,0} + p_{3,0}| + c ( |p_{1,x}+p_{2,x} + p_{3,x}| + |p_{1,y}+p_{2,y}+p_{3,y}| ) }  
\times \\ && \nonumber 
\frac{1}{i p_{4,0} + E_1}   ~~
\frac{1}{i p_{1,0} + E_2 } ~~
\frac{1}{i \lt(p_{1,0} + p_{2,0}\rt) + E_3 } ~~
\frac{1}{i \lt(p_{1,0} + p_{2,0} + p_{3,0}\rt) + E_4 }
\times \\ && \nonumber 
\frac{1}{i (p_{4,0} - p_{1,0}) + E_5 } ~~
\frac{1}{i (p_{4,0} - p_{1,0}-p_{2,0}) + E_6 }  ~~
\frac{1}{i (p_{4,0} - p_{1,0}-p_{2,0}-p_{3,0}) + E_7 },
\label{eq:IG example}
\end{eqnarray} 
where $p_r$ is the set of internal three-momenta, 
and $E_i$ represents the energy of the fermion in the $i$-th fermion propagator as denoted in \fig{fig:HL3}, 
\bqa
E_1 & = & v p_{4,x} + p_{4,y}, \nn
E_2 & = & v p_{1,x} - p_{1,y}, \nn 
E_3 & = & v (p_{1,x} + p_{2,x}) + ( p_{1,y} + p_{2,y}), \nn
E_4 & = &  v (p_{1,x} + p_{2,x} + p_{3,x})   - (p_{1,y} + p_{2,y} + p_{3,y})  , \nn
E_5 & = &  v (- p_{1,x}+p_{4,x} ) - (- p_{1,y}+p_{4,y} ), \nn 
E_6 & = &  v (- p_{1,x} - p_{2,x} + p_{4,x}) +  (- p_{1,y} - p_{2,y} + p_{4,y}), \nn
E_7 & = &  v (- p_{1,x} - p_{2,x} - p_{3,x} + p_{4,x}) - (- p_{1,y} - p_{2,y} - p_{3,y} + p_{4,y}).
\label{E_i}
\eqa

Since frequency integrations are not affected by $v$ and $c$, we focus on the spatial components of momenta from now on.
Our aim is to change the variables for the internal momenta 
so that the enhancement factors of $1/v$ and $1/c$ become manifest. 
As our first three new variables we choose $p_{j}^{'} \equiv c p_{j,x}$ with $1 \leq j \leq 3$.
The last five variables are chosen to be $p_{l+3}^{'} \equiv E_l$ with $1 \leq l \leq 5$.
The transformation between the old variables, written as $\{v p_{i,x}, p_{i,y}\}$, and the new variables is given by
\bqa
\left(
\begin{array}{c}
	p_{1}^{'} \\
	p_{2}^{'} \\
	\vdots \\
	p_{8}^{'} \\
\end{array}
\right)
= 
\left(
\begin{array}{cc}
	\frac{c}{v} {\mathbb I}_{3 }  & 0  \\
	\td {\mathbb A} & \td {\mathbb V}
\end{array}
\right)
\left(
\begin{array}{c}
	v p_{1,x} \\
	v p_{2,x} \\
	v p_{3,x} \\
	v p_{4,x} \\
	p_{1,y} \\
	p_{2,y} \\
	p_{3,y} \\
	p_{4,y} \\
\end{array}
\right),
\label{eq:TRM example}
\eqa
where 
$\td {\mathbb A}$ and $\td {\mathbb{V}}$ are written as
\bqa
\tilde{ \mathbb{A} } =
\lt(
\begin{array}{ccc}
	0 & 0 & 0 \\
	1 & 0 & 0 \\
	1 & 1 & 0 \\
	1 & 1 & 1 \\
	-1 & 0 & 0 \\
\end{array}
\rt),
\hspace{5mm}
\tilde{\mathbb{V}} =
\lt(
\begin{array}{ccccc}
	1 & 0 & 0 & 0 & 1\\
	0 & -1 & 0 & 0 & 0\\
	0 & 1 & 1 & 0 & 0 \\
	0 & -1 & -1 & -1 & 0 \\
	1 & 1 & 0 & 0 & -1 \\
\end{array}
\rt),
\eqa
and ${\mathbb I}_{3 }$ is the $3 \times 3$ identity matrix.
For non-zero $v,c$, 
the change of variables is non-degenerate,
and the Jacobian of the transformation is $(2 c^{3} v)^{-1}$.
We show in the following section that such a non-degenerate choice is always possible for general diagrams. 
An easy mnemonic is that
each fermion loop contributes a factor of $1/v$ 
because of nesting in the small $v$ limit,
while each mixed loop contributes a factor of $1/c$ 
because of the vanishing boson velocity.

In the new coordinates,
the momentum integration in Eq. (\ref{eq:IG example}) becomes
\bqa
I & \sim & \f{v^{3}}{c^3}
\int 
\prod_{i=1}^{8} d p_{i}^{'} 
\left(
\prod_{j=1}^{3}
\frac{1}{ |p_{j}^{'}| + O(c)  } 
\right) 
\left(
\prod_{l=4}^{8} 
\frac{1}{  p_l^{'} }
\right)
\td R[ p^{'} ],
\label{eq:IG3 example}
\eqa 
where $\td R[ p^{'} ]$ includes the propagators that are not explicitly shown.
Now, we can safely take the small $c$ limit inside the integrand,
because every momentum component  has at least one propagator
which guarantees that the integrand decays at least as $1/p_j^{'}$ in the large momentum limit. 
Therefore, the integrations are UV convergent up to potential logarithmic divergences.
To leading order in small $v$, the diagram scales as
\begin{eqnarray*} 
I  \sim  \lt(\f{v}{c}\rt)^3 \sim  v^{\f32}
\end{eqnarray*} 
up to potential logarithmic corrections.

\subsection{General upper bound}

Here we provide a general proof for the upper bound, 
by generalizing the example discussed in the previous section.
We consider a general $L$-loop diagram that includes fermions from patches $1,3$, 
\bqa
I & \sim & v^{\frac{V}{2}} 
\int 
\prod_{r=1}^L d p_r 
\left(
\prod_{l=1}^{I_f} 
\frac{1}{ 
i k_{l,0} + v k_{l,x} + (-1)^{\frac{n_l-1}{2}} k_{l,y} 
}
\right)
\left( 
\prod_{m=1}^{I_b} 
\frac{1}{ |q_{m,0}| + c ( |q_{m,x}| + |q_{m,y}| )   } \right). \hspace{10mm}
\label{eq:IG}
 \eqa 
Here $V$ is the number of vertices. 
$I_f, I_b$ are the numbers of internal fermion and boson propagators, respectively.
$p_r$ is the set of internal three-momenta.
$k_l$ ($q_m$) represents
the momentum that flows through
the $l$-th fermion ($m$-th boson) propagator.
These are linear combinations of
the internal momenta and external momenta.
The way $k_l, q_m$ depend on $p_r$ is determined by 
how we choose internal loops within a diagram.
$n_l=1,3$ is the patch index for the $l$-th fermion propagator.
Since the frequency integrations are not affected by $v$ and $c$,
we focus on the spatial components of momenta from now on.

It is convenient to choose loops
in such a way that 
there exists a propagator 
exclusively assigned to 
each internal momentum.
For this, we follow the procedure given in Sec. VI of \cite{PhysRevB.90.045121}.
For a given diagram, 
we cut internal propagators one by one.
We continue cutting until all loops disappear 
while the diagram remains connected.
First, we cut one fermion propagator in every fermion loop,
which requires cutting $L_f$ fermion lines.
The remaining $L_m \equiv L - L_f$ loops, which we call mixed loops, can be removed by cutting boson propagators.
After cutting $L$ lines  in total, we are left with a connected tree diagram.
Now we glue the propagators back one by one to restore the original $L$-loop diagram.
Every time we glue one propagator, we assign one internal momentum 
such that it goes through the propagator that is just glued back and the connected tree diagram only.
This guarantees that the propagator 
depends only on the internal momentum
which is associated with the loop 
that is just formed by gluing.
In gluing $L_f$ fermion propagators, 
the associated internal momenta 
go through the fermion loops.
The $L_m$ mixed loops necessarily include both fermion and boson propagators. 
After all  propagators are glued back, 
$L$ internal momenta are assigned  
in such a way that for every loop momentum
there is one exclusive propagator.

With this choice of loops, \eq{eq:IG} is written as
\bqa
&& I  \sim  v^{\frac{V}{2}} 
\int 
\prod_{r=1}^L d p_{r,x} dp_{r,y}
\left(
\prod_{j=1}^{L_m}
\frac{1}{ c |p_{j,x}| + c |  p_{j,y} |    } 
\right) 
\left(
\prod_{l=1}^{I_f} 
\frac{1}{ 
E_l ( p ) 
}
\right)
R[ p ].  
\label{eq:IG2}
 \eqa 
Here, frequency is suppressed,
and IR divergences in the integrations over spatial momenta
are understood to be cut off by frequencies.
Our focus is on the UV divergence that arises 
in the spatial momentum integrations
in the limit of small $v$ and $c$.
The first group in the integrand represents the 
exclusive boson propagators assigned to the $L_m$ mixed loops.
Each of the $L_m$ boson propagators
depends on only one internal momentum
due to the exclusive nature of our choice of loops.
The second group represents all fermion propagators.
$E_l(p)$ is the energy of the fermion in the $l$-th fermion propagator
which is given by a linear superposition of $p_{r,x}, p_{r,y}$.
$R[p]$ represents the rest of the boson propagators
that are not assigned as exclusive propagators.

Our strategy is to find a new basis for the loop momenta 
such that the divergences in the small $v$ and $c$ limit become manifest.
The first $L_m$ variables are chosen to be $c p_{j,x}$ with $j=1,2,..,L_m$
while the remaining $2L - L_m$ variables are chosen among $\{ E_l(p) \}$.
This is possible because $I_f \geq (2L - L_m)$ for diagrams with $E>0$.
We express $p_{j}^{'}  \equiv  c p_{j,x}$ and $E_l(p)$ in terms of $v p_{r,x}, p_{r,y}$,
\bqa
\left(
\begin{array}{c}
p_{1}^{'} \\
p_{2}^{'} \\
\vdots \\
p_{L_m}^{'} \\
E_1 \\
E_2 \\
\vdots \\
E_{I_f}
\end{array}
\right)
= 
\left(
\begin{array}{cc}
\frac{c}{v} {\mathbb I}_{L_m }  & 0  \\
 {\mathbb A} & {\mathbb V}
\end{array}
\right)
\left(
\begin{array}{c}
v p_{1,x} \\
v p_{2,x} \\
\vdots \\
v p_{L_m,x} \\
v p_{L_{m}+1,x} \\
\vdots \\
v p_{L,x} \\
p_{1,y} \\
p_{2,y} \\
\vdots \\
p_{L,y} \\
\end{array}
\right).
\eqa
Here ${\mathbb I}_{a }$ is the $a \times a$ identity matrix.
$A_{l,j} = \frac{1}{v} \frac{ \partial E_l }{\partial p_{j,x} }$ 
with $1 \leq l \leq I_f$, $1 \leq j \leq L_m$.
${\mathbb V}$ is an $I_f \times (2L - L_m)$ matrix whose first $L-L_m$ columns
are given by $V_{l,i-L_m} = \frac{1}{v} \frac{ \partial E_l }{\partial p_{i,x} }$ with $L_m+1 \leq i \leq L$
and the remaining $L$ columns are given by
$V_{l,i+(L-L_m)} = \frac{ \partial E_l }{\partial p_{i,y} }$ with $1 \leq i \leq L$.
Now we focus on the lower-right corner of the transformation matrix
which governs the relation between $\vec E^{T} \equiv (E_1, E_2,..,E_{I_f})$
and $\vec P^{T} \equiv ( v p_{L_m+1,x},..,v p_{L,x}, p_{1,y},..,p_{L,y})$
when $p_{j,x}=0$ for $1 \leq j \leq L_m$,
\bqa
\vec E = {\mathbb V} \vec P.
\eqa
$\vec P$ represents
the $x,y$ components of momenta in the fermion loops
and the $y$ components of momenta in the mixed loops.
The matrix ${\mathbb V}$ can be viewed as a collection of $2L-L_m$ column vectors,
each of which have $I_f$ components.
We first show that the $2L-L_m$ column vectors are linearly independent.

 \begin{figure}[h]
 \centering
 \subfigure[]
 {\includegraphics[scale=1.2]{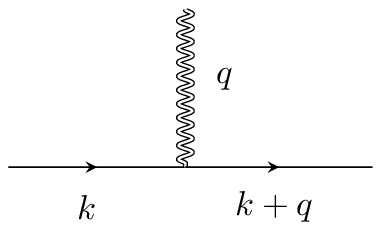}
 \label{fig:EV}} ~~~~~
 \subfigure[]
 {\includegraphics[scale=1.5]{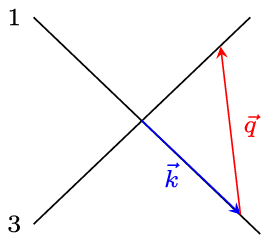}
 \label{fig:EB}} ~~~~~
 \caption{
 For a boson momentum $\vec q$,
 there exists a unique $\vec k$
 such that $\varepsilon_1(\vec k) = \varepsilon_{3}(\vec k + \vec q)=0$
 for $v \neq 0$.
  }
 \label{fig:EVB}
 \end{figure}

If the column vectors were not linearly independent, 
there would exist a nonzero $\vec P$ such that ${\mathbb V} \vec P = 0$.
This implies that there exists at least a one-parameter family of 
$x,y$-momenta in the $L_f$ fermion loops
and $y$-momenta in the $L_m$ mixed loops
such that all internal fermions lie on the Fermi surface.
However, this is impossible for the following reason.
For  $v \neq 0$, a momentum on an external boson leg 
uniquely fixes the internal momenta on the two fermion lines 
attached to the boson line  
if both fermions are required to have zero energy.
This is illustrated in \fig{fig:EVB}.
Similarly, a momentum on an external fermion leg
fixes the momenta on the adjacent internal fermion and boson lines 
if the internal fermion is required to have zero energy 
and only the $y$ component of momentum is allowed to vary in the mixed loops.
Once the momenta on the internal lines attached to the external lines are fixed,
those internal lines in turn fix the momenta of other adjoining internal lines.
As a result, all internal momenta are successively fixed by external momenta
if we require that $E_l=0$ for all $l$.
Therefore, there cannot be a non-trivial $\vec P$ that satisfies ${\mathbb V} \vec P = 0$.
This implies that the column vectors in ${\mathbb V}$ must be linearly independent.

Since ${\mathbb V}$ is made of $(2L-L_m)$ independent column vectors,
it necessarily includes $(2L-L_m)$ independent row vectors.
Let the $l_k$-th rows with 
$k=1,2,..,(2L-L_m)$ 
be the set of rows 
that are linearly independent,
and 
$\td {\mathbb V}$ 
be a $(2L-L_m) \times (2L-L_m)$ invertible matrix
made of these rows.
We choose $p_{L_m+k}^{'} \equiv E_{l_k}$  with
$k=1,2,..,(2L-L_m)$ as the remaining $(2L-L_m)$ integration variables.
The transformation between the original $2L$ momentum variables
and the new variables is given by
\bqa
\left(
\begin{array}{c}
p_{1}^{'} \\
p_{2}^{'} \\
\vdots \\
p_{2L}^{'} \\
\end{array}
\right)
= 
\left(
\begin{array}{cc}
\frac{c}{v} {\mathbb I}_{L_m }  & 0  \\
 \td {\mathbb A} & \td {\mathbb V}
\end{array}
\right)
\left(
\begin{array}{c}
v p_{1,x} \\
v p_{2,x} \\
\vdots \\
v p_{L,x} \\
p_{1,y} \\
p_{2,y} \\
\vdots \\
p_{L,y} \\
\end{array}
\right),
\label{eq:TRM}
\eqa
where 
$\td {\mathbb A}$ is a $(2L-L_m) \times L_m$ matrix
made of the collection of the $l_k$-th rows of ${\mathbb A}$
with $k=1,2,..,(2L-L_m)$.
The Jacobian of the transformation is given by $Y^{-1} c^{-L_m} v^{-L_f}$.
Here, $Y = |\det \td {\mathbb V}|$ is a constant independent of $v$ and $c$,
which is nonzero because $\td {\mathbb V}$ is invertible.

In the new variables, 
\eq{eq:IG2} becomes
\bqa
I & \sim & v^{\frac{V}{2}-L_f}  c^{-L_m}
\int 
\prod_{i=1}^{2 L} d p_{i}^{'} 
\left(
\prod_{j=1}^{L_m}
\frac{1}{ |p_{j}^{'}| + O(c)  } 
\right) 
\left(
\prod_{l=L_m + 1}^{2L} 
\frac{1}{ 
p_l^{'}
}
\right)
\td R[ p^{'} ].
\label{eq:IG3}
 \eqa 
Every component of the loop momenta has at least one propagator
which guarantees that the integrand decays at least as $1/p^{'}_l$ in the large momentum limit.
$\td R[ p^{'} ]$ is the product of all remaining propagators. 
Therefore, the integrations over the new variables are 
convergent up to potentially logarithmic divergences. 
Using $L = \f12 (V + 2 - E)$,
one can see that a general diagram is 
bounded by 
\bqa
I \sim v^{\frac{E-2}{2}} \left(  \frac{v}{c} \right)^{L - L_f}
\eqa
up to logarithmic corrections.
Diagrams with large $(L-L_f)$ are systematically suppressed for $v \ll c$. 
This bound can be checked explicitly for individual diagrams.

\renewcommand\thefigure{B\arabic{figure}}
\renewcommand\thetable{B\arabic{table}}
\renewcommand\theequation{B\arabic{equation}}

\section{Derivation of the self-consistent boson self-energy}
\label{sec:2Loop}

In this section, we derive Eqs. (\ref{eq:D}) and (\ref{eq:cv}) from \eq{eq:SD2}.

The one-loop quantum effective action of the boson generated from Fig. \ref{fig:SEb} is written as
\eqaln{
\Gamma^{1L}_{(0,2)} =  \frac{1}{4} \int dq ~ \Pi^{1L}(q) ~ \text{Tr}\lt[ \Phi(-q) \Phi(q) \rt],
}
where 
\bqa
\Pi^{1L}(q) =
-\pi v  \sum_{n=1}^4
   \int dk ~
   \text{Tr}\lt[\g _1 G_{n}^{(0)}(k) \g _1 G_{\bar{n}}^{(0)}(k+q)\rt]
\eqa
and the bare fermion propagator is 
$G_{n}^{(0)}(k) = -i \f{k_0 \g_0 + \varepsilon_{n}(\vec{k}) \g_1}{k_0^2 + \varepsilon_{n}^2(\vec{k})}$
and $dk \equiv \frac{d^3k}{(2\pi)^3}$.
The integration of the spatial momentum  gives 
$	\Pi^{1L}(q)  =
	 -\f{1}{2}
	\int dk_0\f{(k_0+q_0)k_0}{|k_0 + q_0| |k_0|}$.
The $k_0$ integration generates a linearly divergent mass renormalization 
which is removed by a counter term,
and a finite self-energy,
\eqaln{
	\Pi^{1L} =  |q_0|.
}

Since the one-loop self-energy depends only on frequency,
we have to include higher-loop diagrams to generate a momentum-dependent quantum effective action,
even though they are suppressed by powers of $v$ compared to the one-loop self-energy.
According to \eq{EST}, the next leading diagrams are the ones with $L-L_f = 1$.
Among the diagrams with $L-L_f=1$, 
the only one that contributes to the momentum-dependent boson self-energy is shown in  \fig{fig:2LSEb}.
In particular, other two-loop diagrams that include fermion self-energy insertions do not contribute.
Since the two-loop diagram itself depends on the unknown dressed boson propagator,
we need to solve the self-consistent equation for $D(q)$ in \eq{eq:SD2}.
Here, we first assume that the solution takes the form of \eq{eq:D} with $v\ll c\ll 1$ 
to compute the two-loop contribution,
and show that the resulting boson propagator agrees with the assumed one.
The two-loop self-energy reads
	\begin{align}
	\begin{split}
		\Pi^{2L}(q) = 
-\frac{ \pi^2 v^2 }{2} 
\sum^{4}_{n=1}
\int \dd k \dd p
\left[\frac{1}{\left(k_0+p_0-i \varepsilon _n(\vec{k}+\vec{p})\right)  \left(k_0-i \varepsilon _{\bar{n}}(\vec{k})\right)}\right.\\
		\left.\times\frac{1}{ \left(k_0+q_0-i \varepsilon _n(\vec{k}+\vec{q})\right) \left(k_0+p_0+q_0-i \varepsilon_{\bar{n}}(\vec{k}+\vec{p}+\vec{q})\right)}\right]D(p)+\mathrm{c.c.}.
		\end{split}
	\end{align}
Here c.c. denotes the complex conjugate.
Straightforward integrations over $\vec k$ and $k_0$ give
		\begin{align}\label{eq:provis2}
		\Pi^{2L}(q_0, \vec q) = 
-\frac{\pi v }{8}
\sum^{4}_{n=1}
\int \dd p
\left[\frac{|q_0|-|p_0|}{((p_0+q_0)-i\varepsilon_{\bar{n}}(\vec{p}+\vec{q}))((q_0-p_0)-i\varepsilon_{n}(\vec{q}-\vec{p}))}\right]D(p)+\mathrm{c.c.} .
		\end{align}
Since the frequency-dependent self-energy is already generated from the lower order one-loop graph in \fig{fig:SEb},
we focus on the momentum-dependent part.
This allows us to set the external frequency to zero to rewrite \eq{eq:provis2} as 
		\begin{align}
		\Pi^{2L}(0,\vec{q}) =
\frac{\pi v}{4} 
\sum^{4}_{n=1}
\int \dd p
\left[\frac{|p_0|}{(ip_0+\varepsilon_{\bar{n}}(\vec{p}+\vec{q}))(ip_0+\varepsilon_{n}(\vec{p}-\vec{q}))}\right]D(p).
		\end{align}
		\noindent 
After subtracting the linearly divergent mass renormalization,  
$\Delta \Pi^{2L}(0,\vec{q}) \equiv \Pi^{2L}(0,\vec{q}) - \Pi^{2L}(0,0)$
is UV finite,
			\begin{align}\label{eq:provisional}
			\Delta \Pi^{2L}(0,\vec{q}) = 
\frac{\pi v}{4}
\sum^{4}_{n=1}
\int \dd p 
\frac{|p_0|\mathcal{F}^{1L(n)}(p_0,\vec{p},\vec{q};v)}{(p^2_0+\varepsilon^2_{\bar{n}}(\vec{p}+\vec{q}))(p^2_0+\varepsilon^2_{n}(\vec{p}-\vec{q}))(p^2_0+\varepsilon^2_{\bar{n}}(\vec{p}))(p^2_0+\varepsilon^2_{n}(\vec{p}))}D(p),
			\end{align}
where 
			\begin{align}\label{eq:FunctionF}
			\begin{split}
			\mathcal{F}^{1L(n)}(p_0,\vec{p},\vec{q};v)&=(p^2_0+\varepsilon^{2}_{n}(\vec{p}))(p^2_0+\varepsilon^{2}_{\bar{n}}(\vec{p}))(ip_0-\varepsilon_{\bar{n}}(\vec{p}+\vec{q}))(ip_0-\varepsilon_{n}(\vec{p}-\vec{q}))\\
			&-(p^2_0+\varepsilon^{2}_{\bar{n}}(\vec{p}+\vec{q}))(p^2_0+\varepsilon^{2}_{n}(\vec{p}-\vec{q}))(ip_0-\varepsilon_{\bar{n}}(\vec{p}))(ip_0-\varepsilon_{n}(\vec{p})).
			\end{split}
			\end{align}
Now we consider the contribution of each hot spot separately. 
For $n=1$, the dependence on $q_x$ is suppressed by $v$ compared to the $q_y$-dependent self-energy.
Therefore, we set $q_x=0$ for small $v$.
Furthermore, the $p_y$ dependence in $D(p)$ can be safely dropped 
in the small $c$ limit 
because $\varepsilon_{1}(\vec{p})$ and $\varepsilon_{3}(\vec{p})$
suppress the contributions from large $p_y$.
Rescaling the momentum as $(p_0,p_x,p_y) \rightarrow |q_y| (p_0,p_x/c,p_y)$ followed by the integration over $p_y$,
we obtain the contribution from the hot spot $n=1$,
			\begin{align}
			\Delta \Pi^{2L}(0,\vec{q}) &=
			\frac{v}{32 \pi  c} |q_y|
			\int\dd p_0 \dd p_x \frac{(1+p^2_0-3p^2_x w^2)p^2_0}{(p^2_0+w^2p^2_x)(p^2_0+(w p_x-1)^2)(p^2_0+(wp_x+1)^2)}\frac{1}{|p_0|+|p_x|},
			\end{align}
			where $w \equiv v/c$.
In the integrand, we can not set $w=0$ because the integration over $p_x$ 
is logarithmically divergent in the small $w$ limit,
			\begin{align}
			\begin{split}
&		\Delta \Pi^{2L (1)}(0,\vec{q}) = 
\frac{v}{32 \pi c}   |q_y|
\int\dd p_0 \frac{1}{ 1 + p^2_0 } 
\left[ 
-2\log(w)
-2 p_0 \cot^{-1}(p_0)  
+ p_0^2 \log\left(\frac{p^2_0}{1+p^2_0}\right) 
+ O(w) 
\right].
			\end{split}
			\end{align}
Finally, the integration over $p_0$ gives
			\begin{align}
				\Delta \Pi^{2L (1)}(0,\vec{q}) &=
\frac{|q_y| v}{16 c}
				\left[ \log\left(\frac{1}{w}\right) - 1 + O(w) \right].
			\end{align} 
In the small $w$ limit, the first term dominates.			
Hot spot $3$ generates the same term,
and the contribution from hot spots  $2,4$ is obtained by replacing $q_y$ with $q_x$.
Summing over contributions from all the hot spots, we obtain 
					\begin{align}
					\Delta \Pi^{2L}(0,\vec{q}) & =
\frac{v}{8c}\log\left(\frac{c}{v}\right)\left(|q_x|+|q_y|\right) 
+ O\left( \frac{1}{vc } \right).
					\end{align}

The two-loop diagram indeed reproduces the assumed form of the self-energy 
which is proportional to $|q_x| + |q_y|$ to the leading order in $v$.
The full Schwinger-Dyson equation now boils down to a self-consistent equation for the boson velocity,
 		\begin{align}\label{eq:cvequation}
 		c =\frac{v}{8c} \log\left(\frac{c}{v}\right).
 		\end{align}
$c$ is solved in terms of $v$ as
 		\begin{align}
 		c(v) = \frac{1}{4}\sqrt{v\log\left(\frac{1}{v}\right)}  \left( 1 + O \left( \frac{ \log \log (1/v)}{ \log (1/v)   }  \right)  \right). 
 		\end{align}
This is consistent with the assumption that $v \ll c \ll 1$ in the small $v$ limit.


The full propagator of the boson 
which includes the bare kinetic term in \eq{eq:3D_theory}
is given by
\bqa
D(q)^{-1} = |q_0| + c ( |q_x| + |q_y| ) + \frac{q_0^2}{\tilde \Lambda} + \frac{c_0^2}{\tilde \Lambda} |\vec q|^2,
\eqa
where $\tilde \Lambda$ is a UV scale associated with the coupling.
Depending on the ratio between $c$ and $c_0$, 
which is determined by microscopic details,
one can have different sets of crossovers.

\begin{table}
\centering
 \begin{tabular}{cccc}
\hline
   energy & ~ & scaling & dynamical critical exponent \\
\hline
$q_0   > \tilde \Lambda $ & ~ & $q_0  \sim c_0 q $ & $z=1$ \\
$\frac{c^2}{c_0^2} \tilde \Lambda < q_0   < \tilde \Lambda $ & ~ & $ 
q_0  \sim c_0^2  q^2 / \tilde \Lambda $ & $z=2$ \\
$ q_0   < \frac{c^2}{c_0^2} \tilde \Lambda $ & ~ & $ q_0  \sim c q $ & $z=1$ \\
\hline
 \end{tabular}
\caption{The energy dependent dynamical critical exponent for $c_0 > c$.}
\label{CO1}
\end{table}

\begin{table}
\centering
 \begin{tabular}{cccc}
\hline
   energy & ~ & scaling & dynamical critical exponent \\
\hline
$q_0   > \frac{c}{c_0} \tilde \Lambda $  & ~ & $q_0  \sim c_0 q $ & $z=1$ \\
$\tilde \Lambda < q_0   < \frac{c}{c_0} \tilde \Lambda $ &~& $ q_0  \sim \sqrt{c \tilde \Lambda q}  $ & $z=\frac{1}{2}$ \\
$ q_0  < \tilde \Lambda $ & ~ & $ q_0  \sim c q $ & $z=1$ \\
\hline
 \end{tabular}
\caption{The energy dependent dynamical critical exponent for $c_0 < c$.}
\label{CO2}
\end{table}

For $c_0 > c$, one has a series of crossovers
from the Gaussian scaling with $z=1$ at high energies,
to the scaling with $z=2$ at intermediate energies
and to the non-Fermi liquid scaling with $z=1$ at low energies.
In the low energy limit, the system eventually becomes superconducting.
For $c_0 <  c$, on the other hand, the $z=2$ scaling is replaced
with a scaling with $z=\frac{1}{2}$ at intermediate energies.
This is summarized in Tables \ref{CO1} and \ref{CO2}.

\renewcommand\thefigure{C\arabic{figure}}
\renewcommand\thetable{C\arabic{table}}
\renewcommand\theequation{C\arabic{equation}}

\section{Derivation of the beta function for $v$}
\label{sec:1L}

In this section, we derive the beta function for $v$ in \eq{eq:beta}.
We first compute the counter terms 
that need to be added to the local action 
such that the quantum effective action
is independent of the UV cut-off scale
to the lowest order in $v$.
Then we derive the beta function for $v$
and its solution, 
which confirms that 
$v$ flows to zero in the low-energy limit.

\subsection{Frequency-dependent fermion self-energy}

\begin{figure}[!ht]
\centering
\includegraphics[scale=1.1]{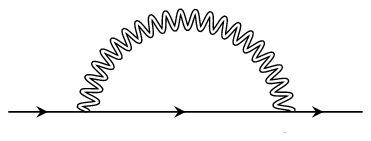}
\caption{
The one-loop diagram for the fermion self-energy.
}
\label{fig:SEf}
\end{figure}

According to \eq{EST}, 
the leading order fermion self-energy 
is generated from Fig. \ref{fig:SEf}
in the small $v$ limit.
The one-loop fermion self-energy for patch $n$ is given by 
\eqaln{
\Sigma^{1L(n)}(k_0,\vec k) =
\f{3 \pi v}{2} 
   \int dp ~ 
   \g_{1} G_{\bar{n}}^{(0)}(p+k) \g_{1} D(p),
   \label{eq:fermion self energy}
   }
where the dressed boson propagator is 
$ D(p) = \f{1}{|p_0| + c(v) (|p_x| + |p_y|)}$.
We first compute  $\Sigma^{1L(n)}(k)$ for  $n= 1$.
The quantum correction is logarithmically divergent,
and a UV cut-off $\Lambda$ is imposed 
on $p_y$,
which is the momentum perpendicular to the Fermi surface 
for $n=1$ in the small $v$ limit.
However, the logarithmically divergent term is independent of 
how UV cut-off is implemented.
To extract the frequency-dependent self-energy, we set $\vec k=0$
and rescale 
$(p_0, p_x, p_y ) \rightarrow |k_0| ( p_0, p_x/c, p_y )$  to rewrite
\eqaln{
	\Sigma^{1L(1)}(k_0,0)
	& =
	i \g_0 k_0 \f{3 \pi v}{ 2 c}
	\int dp ~ 
	\f{p_0 + 1}
	{\lt[
		(p_0 + 1)^2 + (w p_x - p_y)^2
		\rt]
		\lt[
		|p_0| + |p_x| + c |p_y|
		\rt]},
}
where $w = \f{v}{c}$.
Under this rescaling, the UV cut-off for $p_y$ is also rescaled to $\Lambda_0 = \Lambda/|k_0|$. 
The $p_0$ integration gives
\bqa
&	\Sigma^{1L(1)}(k_0,0)    =
	i \g_0 k_0 \f{3 \pi v}{ 2 (2\pi)^3  c }
	\int_{-\Lambda_0}^{\Lambda_0} dp_y  \int dp_x \nn 
& 	\lt[\f{\pi}{2} |p_y - w p_x|
	\lt(\f{1}{(p_y - w p_x)^2 + (-1 + |p_x| + c |p_y|)^2} - \f{1}{(p_y - w p_x)^2 + (1 + |p_x| + c |p_y|)^2}\rt)
	\rt. \nn 
& 	- \lt. 
	(p_y - w p_x) \arccot(p_y - w p_x)
	\lt(\f{1}{(p_y - w p_x)^2 + (-1 + |p_x| + c |p_y|)^2} + \f{1}{(p_y - w p_x)^2 + (1 + |p_x| + c |p_y|)^2}\rt) 
	\rt. \nn
& + 	\lt.
	\f12 \log\lt(\f{1 + (p_y - w p_x)^2}{(|p_x| + c |p_y|)^2}\rt)
	\lt(\f{1 + |p_x| + c |p_y|}{(p_y - w p_x)^2 + (1 + |p_x| + c |p_y|)^2} - 
	\f{-1 + |p_x| + c |p_y|}{(p_y - w p_x)^2 + (-1 + |p_x| + c |p_y|)^2}\rt)\rt].
\label{eq:Z_1 post p_0 integration}
\eqa
The logarithmically divergent contribution is obtained to be
\eqaln{
	\Sigma^{1L(1)}(k_0,0)
	& =
	\f{3}{4 \pi} \f{v}{c}
	\log\lt(\f{\Lambda}{|k_0|}\rt) i \g_0 k_0
}
in the small $v$ limit.
The self-energy for other patches is obtained from a series of $90$-degree rotations,
and the frequency-dependent part is identical for all patches.
In order to remove the cut-off dependence in the quantum effective action,
we add the counter term, 
\bqa 
\sum_{n=1}^4 \sum_{\s=\uparrow,\downarrow} \int dk ~
\bar{\Psi}_{n,\s}(k) \lt( Z_{1,1} \,  i \g_0 k_0 \rt) \Psi_{n,\s}(k)
\label{eq:CT1}
\eqa
with 
\bqa
Z_{1,1} = -\f{3}{4\pi} \f{v}{c} \, \log\lt(\f{\Lambda}{\mu}\rt),
\label{eq:Z1}
\eqa
where $\mu$ is the scale at which the quantum effective action
is defined in terms of the renormalized velocity $v$.
The counter term guarantees that the renormalized propagator
at the scale $\mu$ is expressed solely in terms of $v$
in the $\Lambda/\mu \rightarrow \infty$ limit.

\subsection{Momentum-dependent fermion self-energy}
To compute the momentum-dependent fermion self-energy, 
we start with Eq. (\ref{eq:fermion self energy}) for $n=1$ and set $k_0 = 0$.
Rescaling $p_x \rightarrow \f{p_x}{c}$ gives
\eqaln{
	\Sigma^{1L(1)}(0,\vec{k})
	=
	-\f{3 \pi v }{2 c} i \g_{1} \int dp ~  
	\f{w p_x - p_y + \varepsilon_{3}(\vec{k})}
	{\lt[p_0^2 + (w p_x - p_y + \varepsilon_{3}(\vec{k}))^2\rt]
		\lt[|p_0| + |p_x| + c|p_y|\rt]}.
}
The integration over $p_0$ results in 
$	\Sigma^{1L(1)}(0,\vec{k})
	= 		\lt. \Sigma^{1L(1)}(\vec{k}) \rt|_{\text{term 1}} + \lt. \Sigma^{1L(1)}(\vec{k}) \rt|_{\text{term 2}}$,
where
\eqaln{
	\lt. \Sigma^{1L(1)}(\vec{k}) \rt|_{\text{term 1}} =
	&
	-i \g_{1} \f{3 \pi v}{2  (2\pi)^3 c} 
		\int dp_y  \int dp_x
	\f{\mathrm{sgn}(w p_x - p_y + \varepsilon_{3}(\vec{k})) (|p_x| + c|p_y|) \pi}
	{(p_y - \varepsilon_{3}(\vec{k}) - w p_x)^2 + (|p_x| + c|p_y|)^2},
	\\ 
	 \lt. \Sigma^{1L(1)}(\vec{k}) \rt|_{\text{term 2}} =
	&
	- i \g_{1} \f{3 \pi v }{2 (2\pi)^3 c }
	\int dp_y  \int dp_x
	\f{(p_y - \varepsilon_{3}(\vec{k}) - w p_x) \log\lt(\f{(|p_x| + c|p_y|)^2}{(p_y - \varepsilon_{3}(\vec{k}) - w p_x)^2}\rt)}
	{(p_y - \varepsilon_{3}(\vec{k}) - w p_x)^2 + (|p_x| + c|p_y|)^2}.
}
We first compute the first term.
After performing the $p_x$ integration, 
we rescale $p_y \rightarrow |\varepsilon_{3}(\vec{k})| p_y$ to obtain
\bqa
	& \lt. \Sigma^{1L(1)}(\vec{k}) \rt|_{\text{term 1}} 
	=
	-\f{3 \pi^2 v}{2  (2\pi)^3 c } 
i \g_{1} \varepsilon_{3}(\vec{k})
	\int_{-\Lambda_3}^{\Lambda_3} dp_y 
	\Bigg[ 
	\f{\pi w}{2 (1 + w^2)} 
	\lt( \mathrm{sgn}\lt(p_y - 1 + c w |p_y|\rt) + 
	\mathrm{sgn}\lt(p_y - 1 - c w |p_y|\rt) \rt)
	\nn &  +
	\f{\mathrm{sgn}\lt(p_y - 1\rt)}{1 + w^2}   
	\lt(w \arctan\lt(\f{w \lt(-p_y + 1\rt) + c |p_y|}{p_y - 1 + c w |p_y|}\rt) + 
	w \arctan\lt(\f{w \lt(p_y - 1\rt) + c |p_y|}{-p_y + 1 + c w |p_y|}\rt) 
	\rt.
	\nn & -
	\lt. 
	2 w \arctan\lt(w^{-1}\rt) 
	- \log\lt(\f{c^2 w^2 p_y^2 + \lt(p_y - 1\rt)^2 + 
		2 c w \lt|p_y - 1\rt| |p_y|}{w^2 \lt(c^2 p_y^2 + \lt(p_y - 1\rt)^2\rt)}\rt)\rt)\Bigg],
\eqa
where $\Lambda_3 = \f{\Lambda}{|\varepsilon_{3}(\vec{k})|}$.
The remaining $p_y$ integration gives
\eqaln{
	\lt. \Sigma^{1L(1)}(\vec{k}) \rt|_{\text{term 1}} 
	= \f{3v(w - c)}{4\pi} \log\lt(\f{\Lambda}{|\varepsilon_{3}(\vec{k})|}\rt) i \g_{1} \varepsilon_{3}(\vec{k}) 
}
to the leading order in $v$
up to terms that are finite in the large $\Lambda$ limit.

The second term can be computed similarly in the small $v$ limit,
\eqaln{
	& \lt. \Sigma^{1L(1)}(\vec{k}) \rt|_{\text{term 2}}
	= -\f{3  }{2 \pi^2} \, v \log\lt(\f1c\rt) \log\lt(\f{\Lambda}{|\varepsilon_{3}(\vec{k})|}\rt) i \g_{1}  \varepsilon_{3}(\vec{k})
\label{Upsilon_20_1}
}
up to UV-finite terms.
It is noted that the second term is dominant for small $v$.

\begin{figure}[!ht]
\centering
\subfigure[]
{\includegraphics[scale=1.1]{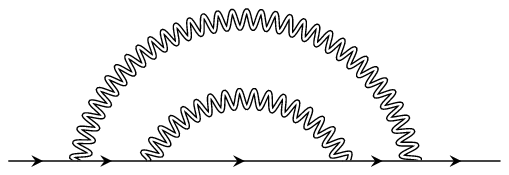}
\label{fig:2LFSE1}} ~~~~~
\subfigure[]
{\includegraphics[scale=1.1]{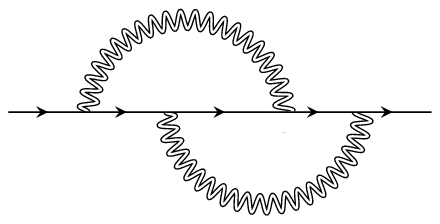}
\label{fig:2LFSE2}} \\
\caption{
Two-loop diagrams for the fermion self-energy.
While (a) is sub-leading in the small $v$ limit,
(b) is of the same order as \fig{fig:SEf}.
}
\label{fig: CT3}
\end{figure}

According to \eq{EST}, the upper bound for the one-loop fermion self-energy is $v/c$.
However,   \eq{Upsilon_20_1} is strictly smaller than the upper bound.
The extra suppression by $c$ arises due to the fact that 
the external momentum in \fig{fig:SEf}
can be directed to flow only through the boson propagator,
and the diagram becomes independent of the external momentum in the small $c$ limit.
Since this suppression does not happen for higher-loop diagrams in general,
the one-loop diagram becomes comparable to some two-loop diagrams with $L-L_f = 2$.
Therefore, we have to include the two-loop diagrams for the self-energy
in order to capture all leading order corrections.
The rainbow diagram in \fig{fig:2LFSE1} is smaller
for the same reason as the one-loop diagram.
Three and higher-loop diagrams remain negligible,
and only \fig{fig:2LFSE2} contributes to the leading order.
The two-loop self-energy for patch $n$ is given by
\eqaln{
\Sigma^{2L(n)}(k_0, \vec k) = 
\frac{ 3 \pi^2 v^2}{4} \int dp dq ~
\left[
\gamma_1
G_{\bar n} (k + q)
\gamma_1
G_{n} (k + q + p)
\gamma_1
G_{\bar n} (k + p)
\gamma_1
\right]
D(q) D(p).
}
It is noted that $\Sigma^{2L(n)}(k_0,0)$ is strictly smaller than $\Sigma^{1L(n)}(k_0,0)$,
and only $\Sigma^{2L(n)}(0,\vec k)$ is of the same order as 
$\Sigma^{1L(n)}(0, \vec k)$.
Therefore, we only compute $\Sigma^{2L(n)}(0, \vec k)$.
After performing the integrations over $p_y,q_y$,
the self-energy for patch $1$ becomes
\eqaln{
	\Sigma^{2L(1)}(0, \vec k) = & 
	- \frac{3 v^2}{2^8 \pi^2 c^2} i \gamma_1
	\int dp_0  \int dq_0
	~ ( \mathrm{sgn}(p_0) + \mathrm{sgn}(p_0 + q_0)) (\mathrm{sgn}(q_0) + \mathrm{sgn}(2 p_0 + q_0)) \times	\nn 
	& \int dp_x \int dq_x ~
	\frac{2w(p_x + q_x) + (3vk_x - k_y)}{4(p_0 + q_0)^2 + (2w(p_x + q_x) + (3vk_x - k_y))^2}
	\frac{1}{|p_0|+|p_x|}
	\frac{1}{|q_0|+|q_x|}.
}
We single out the factor of $(3vk_x - k_y)$ by rescaling $(p_0,p_x,q_0,q_x) \rightarrow |3vk_x - k_y| (p_0,p_x,q_0,q_x)$.
To perform the $p_x$ and $q_x$ integrals,
we introduce variables 
$a = \frac{1}{2} (p_x + q_x), b = \frac{1}{2} (p_x - q_x)$. 
After the straightforward integration over $b$, we rescale $a \rightarrow \frac{a}{w}$ to obtain
\eqaln{
	 \Sigma^{2L(1)}(0, \vec k) & = 
	- \frac{3 v^2}{2^7 \pi^2 c^2} i  \gamma_1(3vk_x - k_y)
	\int  dp_0 \int dq_0 \nn
	& (\mathrm{sgn}(p_0) + \mathrm{sgn}(p_0 + q_0)) (\mathrm{sgn}(q_0) + \mathrm{sgn}(2 p_0 + q_0)) 
	 \int da ~ 
		\frac{4a + 1}{4(p_0 + q_0)^2 + (4a + 1 )^2}  \times \nn
		&	
		 \left(\frac{\log\left(\frac{(2 |a| + w |p_0|) (2 |a| + w |q_0|)}{w^2 |p_0| |q_0|}\right)}{2 |a| + w (|p_0| + |q_0|)}
	- \frac{\log\left(\frac{w |q_0|}{2 |a| + w |p_0|}\right)}{2 |a| + w (|p_0| - |q_0|)} 
	- \frac{\log\left(\frac{w |p_0|}{2 |a| + w |q_0|}\right)}{2 |a| - w (|p_0| - |q_0|)}\right),
}
where the frequency integrations are understood to have a UV cut-off, 
$\Lambda_3^{'} = \frac{\Lambda}{|3vk_x - k_y|}$ in the rescaled variable.
In the small $w$ limit, the $a$ integration diverges as $(\log(w))^2$.
The sub-leading terms are suppressed compared to the one-loop diagram,
and we drop them in the small $w$ limit. 
The remaining frequency integrations are logarithmically divergent in the UV cut-off,
\eqaln{
\Sigma^{2L(1)}(0, \vec k)=
- i \gamma_1 \frac{3}{32 \pi^2} 
\left( \frac{v}{c} \log \frac{c}{v} \right)^2 
\log \lt( \frac{\Lambda}{ | 3 v k_x - k_y|} \rt)
\left( 3 v k_x - k_y \right).
\label{eq:dS_(2,0) full_2}
}
This is of the same order as \eq{Upsilon_20_1}
because of $\left( \frac{v}{c} \log \frac{c}{v} \right)^2  = 8 v \log \frac{1}{c}$ 
to the leading order in $v$.

The vertex correction in \fig{fig:2LFSE2} strengthens the bare vertex,
and the two-loop self-energy has the same sign as the one-loop self-energy.
In particular, both the one-loop and two-loop quantum corrections enhance nesting, 
and drive $v$ to a smaller value at low energies.
To remove the cut-off dependences of \eq{Upsilon_20_1} and \eq{eq:dS_(2,0) full_2}
in the quantum effective action,
we add the counter term 
\bqa 
\sum_{\s=\uparrow,\downarrow}
	\int dk ~ 
	\bar{\Psi}_{1,\s}(k) \lt( i \g_1 (Z_{2,1} v k_x + Z_{3,1} k_y) \rt) \Psi_{1,\s}(k)
\label{eq:CT23}
\eqa
	with 
\bqa
Z_{2,1} &=&  \f{15}{4\pi^2} v \, \log\lt(\f1c \rt) \, \log\lt( \f{\Lambda}{\mu} \rt), \nn
Z_{3,1} &=&  -\f{9}{4\pi^2} v \, \log\lt(\f1c \rt) \, \log\lt( \f{\Lambda}{\mu} \rt).
\label{eq:Z23}
\eqa
Counter terms for $n=2,3,4$ are fixed by the four-fold rotational symmetry.

\subsection{Vertex correction}

\begin{figure}[!ht]
\centering
\includegraphics[scale=1.1]{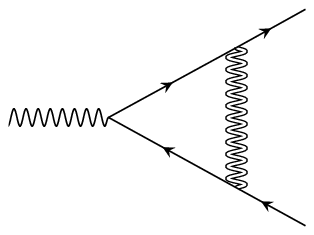}
\caption{
The one-loop diagram for the vertex correction.
}
\label{fig:Y}
\end{figure}

The one-loop vertex correction in Fig. \ref{fig:Y} is given by
\eqaln{
\Gamma^{1L}(k,q) &=
\f{\pi v}{2}
	\int dp ~ \g_1 G_{\bar{n}}^{(0)}(p+k+q) \g_1 G_{n}^{(0)}(p+k) \g_1 D(p).
	\label{eq:vertex correction}
}
We set all external momenta to zero except for $k_0$,
which plays the role of an IR regulator.
For $n=1$, it is convenient to rescale  
$(p_0, p_x, p_y ) \rightarrow |k_0| ( p_0, p_x/c, p_y )$.
The $p_0$ integration gives
\eqaln{
	& \Gamma^{1L(1)}(k_0)  =
	\f{\pi v}{ 2 c} \g_1 \f{1}{(2\pi)^3} \int_{-\Lambda_0}^{\Lambda_0} dp_y \int dp_x
	\nn &
	\f{
		 \lt \{ 
	      \begin{aligned}
			&\lt[\lt((p_y - w p_x) (p_y + w p_x)^3 + (-1 + (|p_x| + c |p_y|)^2)^2 \rt)\lt(1 + (p_y - w p_x)^2 + (|p_x| + c |p_y|)^2\rt) \rt. + \\ &
			\lt. \lt.  (p_y - w p_x) (p_y + w p_x) \lt(1 + 6 (|p_x| + c |p_y|)^2 + (|p_x| + c |p_y|)^4 + (p_y - w p_x)^2 (1 + (|p_x| + c |p_y|)^2)\rt) \rt. \rt. \nn &
			\lt. + (p_y + w p_x)^2 \lt((-1 + (|p_x| + c |p_y|)^2)^2 + (p_y - w p_x)^2 (1 + (|p_x| + c |p_y|)^2)\rt)\rt]
			\log(|p_x| + c |p_y|) 
		\end{aligned}
	    \rt \}
	    }
	    {
		-\frac{1}{2} \lt \{ 
		\begin{aligned}
				& \lt((p_y - w p_x)^2 + (-1 + |p_x| + c |p_y|)^2\rt) \lt((p_y + w p_x)^2 + (-1 + |p_x| + c |p_y|)^2\rt) \nn &
				\times \lt((p_y - w p_x)^2 + (1 + |p_x| + c |p_y|)^2\rt) \lt((p_y + w p_x)^2 + (1 + |p_x| + c |p_y|)^2\rt)
		\end{aligned}
		\rt \}
	    } \nn & +
	\f{
		\lt \{ 
		\begin{aligned}
					&2 \arccot(p_y + w p_x) \lt(1 + (p_y + w p_x)^2 - (|p_x| + c |p_y|)^2\rt)  \nn
				& \lt.	+ (p_y + w p_x) \log(1 + (p_y + w p_x)^2) (1 + (p_y + w p_x)^2 + (|p_x| + c |p_y|)^2) 
					\rt. \nn &
					+ \; \pi \; \mathrm{sgn}(p_y + w p_x) (|p_x| + c |p_y|) (-1 + (p_y + w p_x)^2 + (|p_x| + c |p_y|)^2)
		\end{aligned}
		\rt \}
	  }
	  {
			    2 p_y \lt((p_y + w p_x)^2 + (-1 + |p_x| + c |p_y|)^2\rt) \lt((p_y + w p_x)^2 + (1 + |p_x| + c |p_y|)^2\rt)
	  } \nn & +
	\f{
		\lt \{ 
		\begin{aligned}
						&2 \arccot(p_y - p_x w) (1 + (p_y - w p_x)^2 - (|p_x| + c |p_y|)^2)  \nn
						& \lt. + (p_y - w p_x) \log(1 + (p_y - w p_x)^2) (1 + (p_y - w p_x)^2 + (|p_x| + c |p_y|)^2)  \rt. \nn &
						+ \; \pi \; \mathrm{sgn}(p_y - w p_x) (|p_x| + c |p_y|) (-1 + (p_y - w p_x)^2 + (|p_x| + c |p_y|)^2) 
		\end{aligned}
		\rt \}
	  }
	  {
					2 p_y \lt((p_y - w p_x)^2 + (-1 + |p_x| + c |p_y|)^2\rt) \lt((p_y - w p_x)^2 + (1 + |p_x| + c |p_y|)^2\rt)
	  },
}
where the rescaled cut-off for $p_y$ is $\Lambda_0 = \f{\Lambda}{|k_0|}$. 
After the $\vec p$ integration, the logarithmically divergent contribution is obtained to be
\eqaln{
& \Gamma^{1L(1)}(k_0)  =
\f{1}{4 \pi} 
\f{v}{c} \log\lt(\f{c}{v}\rt)\log\lt(\f{\Lambda}{|k_0|}\rt) \g_1
}
in the small $v$ limit.
The vertex corrections for different $n$ are the same.
The counter term for the vertex  becomes
\bqa
		 Z_{6,1} ~ i \sqrt{\f{\pi v}{2}} \sum_{n=1}^4 \sum_{\s,\s' = \uparrow, \downarrow}
		\int dk \int dq ~ \bar{\Psi}_{n,\s}(k+q) \Phi_{\s,\s'}(q)  \g_1
		\Psi_{\bar{n},\s'}(k)		
\label{eq:CT6}
\eqa
with 
\bqa 
Z_{6,1} = - \f{1}{4 \pi} \f{v}{c} \log \lt( \f{c}{v} \rt) \log\lt(\f{\Lambda}{\mu}\rt). 
\label{eq:Z6}
\eqa
We explicitly check that two-loop vertex corrections
are sub-leading in $v$, in agreement with \eq{EST}.

\subsection{The beta function for $v$}
\label{betav}

The counter terms in Eqs. 
(\ref{eq:CT1}), (\ref{eq:CT23}), (\ref{eq:CT6})
are added to the action in Eq. (\ref{eq:min_theory})
to obtain the bare action,
\eqaln{
	\mathcal{S}_B & = 
	\sum_{n=1}^4 
	\sum_{\sigma=\uparrow, \downarrow}
	\int 
	dk~
	\bar{\Psi}_{n,\sigma}(k) 
	\Bigl[ i Z_1 \gamma_0 k_0 + i \gamma_{1} \varepsilon_n^B(\vec{k}) \Bigr] 
	\Psi_{n,\sigma}(k) \nn
& + 
	i Z_6 \, \sqrt{\f{\pi v}{2}}  \sum_{n=1}^4
	\sum_{\sigma,\sigma'}
	\int  dk \; dq ~
	\Bigl[
	\bar{\Psi}_{\bar n,\sigma}(k+q) 
	\Phi_{\sigma,\sigma'}(q)
	\gamma_{1} 
	\Psi_{n,\sigma'} (k) 
	\Bigr],
	\label{eq:bare_theory_app}
}
where 
$\varepsilon_1^B(\vec k) = Z_2 v k_x + Z_3 k_y$,
$\varepsilon_2^B(\vec k) = - Z_3 k_x + Z_2 v k_y$,
$\varepsilon_3^B(\vec k) = Z_2 v k_x - Z_3 k_y$,
$\varepsilon_4^B(\vec k) = Z_3 k_x + Z_2 v k_y$.
Here $Z_n = 1 + Z_{n,1}$ is given in Eqs. (\ref{eq:Z1}), (\ref{eq:Z23}) and (\ref{eq:Z6}).
The bare action generates the physical quantum effective action 
which is expressed solely in terms of the renormalized coupling $v$ 
measured at an energy scale $\mu$.
The relationship between the renormalized and bare quantities is given by
\eqaln{
	k_{x,B} = k_{x}; ~
	k_{y,B} = k_{y}; ~
	k_{0,B} = \frac{Z_1}{Z_3} k_0; ~ 
	v_{B} = \frac{Z_2}{Z_3}v; ~
	\Psi_{B}(k_B) = \frac{Z_3}{Z_1^{\frac{1}{2}}}\Psi(k); ~ 
	\Phi_{B}(k_B) = \f{Z_3^{\frac{1}{2}} Z_6}{Z_1 Z_2^{\frac{1}{2}}} \, \Phi(k).
\label{eq:Bare}
}

The beta function for $v$ is obtained by requiring that the bare coupling $v_B$ does not depend on $\mu$,
\eqaln{
	\lt(Z_2 Z_3 + v \lt(\f{\p Z_2}{\p v} Z_3 - Z_2 \f{\p Z_3}{\p v}\rt)\rt) \beta_v + v \lt(\f{\p Z_2}{\p \log \mu} Z_3 - Z_2 \f{\p Z_3}{\p \log \mu}\rt) = 0.
	\label{eq:beta_v general Z_2}
} 
This gives the beta function 
which describes the flow of $v$ under the change of the scale $\mu$,
\eqaln{
	\f{d v}{d \log \mu} =  \f{6}{\pi^2} \, v^{2} \log\lt[ 4 \left( \frac{1}{ v \log 1/v } \right)^{\f12}  \rt]
} 
to the leading order in $v$.
Introducing a logarithmic scale $\ell = -\log \mu$, 
the beta function can be rewritten as $	\f{d v}{d \ell} =  \f{3}{ \pi^2} \, v^{2} \log v $
up to $\log \log v$.
The solution is given by
\bqa
Ei[ \log 1/v(\ell) ] = Ei[ \log 1/v(0) ] + \f{3}{ \pi^2} \ell,
\eqa
where $Ei(x)$ is the exponential integral function,
which goes as $Ei(x) = e^x \left[ \frac{1}{x} + O( 1/x^2 ) \right] $ in the large $x$ limit.
Therefore, $v$ flows to zero as
\bqa
v(\ell) = \frac{\pi^2}{3} \frac{1}{ \ell ~ \log \ell }
\label{eq:vl}
\eqa
for $\ell \gg \frac{1}{v(0) \log 1/v(0)}$.
For sufficiently large $\ell$, $v(\ell)$ decays to zero in a manner
which is independent of its initial value.
The velocity of the collective mode flows to zero at a slower rate,
\bqa
c(\ell) = \frac{ \pi}{ 4 \sqrt{3}} \frac{1}{ \sqrt{\ell} },
\label{eq:cl}
\eqa
and the ratio $w= v/c$ flows to zero as
\bqa
w(\ell) = \frac{ 4 \pi}{\sqrt{3}} \frac{1}{ \sqrt{\ell}  \log \ell}.
\label{eq:wl}
\eqa

Similarly, 
the multiplicative renormalization for the frequency and fields in \eq{eq:Bare}
generates the deviation of the dynamical critical exponent from one 
and the anomalous dimensions for the fields,
\eqaln{
	\eta_\phi &= \f{d}{d \, \log \mu} \log \lt(\f{Z_3^{\f12} Z_6}{Z_1 Z_2^{\f12}}\rt),
	\\
	\eta_\psi &= \f{d}{d \, \log \mu} \log \lt( \f{Z_3}{Z_1^{\f12}} \rt),
	\\
	z &= 1 + \f{d}{d \, \log \mu} \log \lt(\f{Z_1}{Z_3}\rt)
	\label{eq:critical exponents general Z_n}
}
which reduce to the expressions in Eqs. (\ref{eq:exponents})
to the leading order in $v$.

\renewcommand\thefigure{D\arabic{figure}}
\renewcommand\thetable{D\arabic{table}}
\renewcommand\theequation{D\arabic{equation}}

\section{Derivation of the scaling forms for physical observables}\label{sec:PP}

In this section, 
we derive the expressions for
the Green's functions 
and the specific heat
in Eqs. (\ref{eq:GR}),
(\ref{eq:DR})
and (\ref{eq:c}).

	\subsection{The Green's function}
	\label{sec:SF}
We derive the form of the electron Green's function near hot spot $1+$.
The Green's functions for all other hot spots are determined from that of $1+$ by symmetry.
The  Green's function satisfies the renormalization group equation, 
\bqa
\left[
\frac{1 - 2 \eta_\psi - (z-1)}{z} +  k_0 \frac{\partial}{\partial k_0}
+ \frac{1}{z} \vec k \cdot \frac{ \partial}{\partial \vec k} 
- \frac{\beta_v}{z} \frac{ \partial}{\partial v}
\right] G_{1+}(k_0, \vec k;  v) = 0.
\label{eq:G}
\eqa
The solution becomes
\bqa
G_{1+}(k_0, \vec k; v ) = 
e^{ \int_0^l \frac{ 1 - 2 \eta_\psi( v(l^{'}) ) - [  z( v(l^{'})) -1  ]   }{ z( v(l^{'})  ) } dl^{'} } 
G_{1+}\left( e^l k_0, 
e^{ \int_0^l \frac{1}{ z( v(l^{'})  ) } dl^{'} } \vec k
;  v(l) \right),
\eqa
where $v(l)$ satisfies $\frac{d v(l)}{d l} = - \frac{ \beta_v }{ z( v ) }$ 
with the initial condition $v(0)=v$,
and $z(v)$ and $\eta_\psi(v)$ depend on $l$ through $v(l)$.
We write $\frac{1 - 2 \eta_\psi - (z-1)}{z} = \frac{1}{z}  - 2 \tilde \eta_\psi$,
where $\tilde \eta_{\psi} = \frac{1}{2} \frac{ \partial \log Z_3}{\partial \log \mu}$
to the leading order in $v$.
Although $\tilde \eta_\psi$ is sub-leading compared to $1/z$, 
we keep it because only $\tilde \eta_\psi$ contributes
to the net anomalous dimension of the propagator.
From Eqs. (\ref{eq:vl})-(\ref{eq:wl}),
one obtains the solution to the scaling equation,
	\begin{align}\label{eq:2PointFunction}
	G_{1+}(k_0,\vec{k};v) &=
	\exp\left(l- 2 \sqrt{3}  \frac{\sqrt{l}}{\log(l)} -\frac{3}{8} \log l    \right)
	G_{1+} \left( e^{l}k_0, \exp\left(l- 2\sqrt{3} \frac{\sqrt{l}}{\log(l)}\right)\vec{k},
	\frac{\pi^2}{3}\frac{1}{l\log(l)}\right)
	\end{align}
in the large $l$ limit.
We choose $l=\log(1/k_0)$  
and take the small $k_0>0$ limit with
$\exp\left(l- 2\sqrt{3} \frac{\sqrt{l}}{\log(l)}\right)\vec{k} \sim 1$.
By using the fact that the Green's function is given by  
$G_{1+}(k_0,\vec k; v) = \left( i k_0 + v k_x + k_y \right)^{-1}$ in the small $v$ limit, 
we readily obtain  
\bqa
G_{1+}(k_0,\vec{k};v) &=
\frac{1}{ 
F_\psi(k_0 )
\left[  
i k_0 ~ F_z(k_0)  
+ 
\left(
\frac{ \pi^2 }{3}  \frac{k_x}{  \log \frac{1}{k_0} ~ \log \log \frac{1}{k_0}  }
+ k_y
\right)
\right]
}
\eqa 
in the low-energy limit with fixed $\frac{\vec k}{k_0 F_z(k_0)}$, 
where
$F_{\psi}(k_0 ) = 
\left(
\log \frac{1}{k_0}
\right)^{ \frac{3}{8} }
$ 
and
$
F_z(k_0) = e^{  2 \sqrt{3} \frac{ \left( \log \frac{1}{k_0}  \right)^{1/2}}{ \log \log \frac{1}{k_0} }}
$.
The analytic continuation to the real frequency gives \eq{eq:GR}.

Similarly, the Green's function of the boson satisfies
\bqa
\left[
\frac{1 - 2 \eta_\phi - (z-1)}{z} +  q_0 \frac{\partial}{\partial q_0}
+ \frac{1}{z} \vec q \cdot \frac{ \partial}{\partial \vec q} 
- \frac{\beta_c}{z} \frac{ \partial}{\partial c}
\right]D(q_0, \vec q;  c) = 0,
\label{eq:RGD}
\eqa
where $\beta_c = \frac{ d c}{d \log \mu}$.
Here we view the boson propagator as a function of $c$ instead of $v$
because it depends on $v$ only through $c$ to the leading order.
However, this does not affect any physical observable 
since in the end there is only one independent parameter. 
The solution to the scaling equation takes the form,
	\begin{align}
	D(q_0,\vec{q},c) &= \exp\left(l -\frac{2 \sqrt{l}}{\sqrt{3}}  -2 \sqrt{3}\frac{\sqrt{l}}{\log l}    \right)
	D\left( e^{l} q_0,\exp\left(l-2\sqrt{3}\frac{\sqrt{l}}{\log(l)}\right)\vec{q};
	\frac{ \pi}{4 \sqrt{3}} \frac{1}{ \sqrt{l} }
	 \right).
	\end{align}
By choosing $l=\log(1/q_0)$ 
and using the fact that the boson propagator is given by \eq{eq:D} in the limit of small $v$ and $c$, 
we obtain
\bqa
D(q_0, \vec q) =  \frac{1}{
F_{\phi} (q_0)
\left( 
|q_0| F_z(q_0)
+
\frac{ \pi}{4 \sqrt{3}}
\frac{ |q_x| + |q_y|  }
{\left( \log \frac{1}{q_0} \right)^{1/2}  }
\right)
}
\eqa
in the low-energy limit with fixed $\frac{\vec q}{q_0 F_z(q_0)}$.
Here
$F_{\phi} (q_0)
\equiv e^{ \frac{2}{\sqrt{3}} \left( \log \frac{1}{ q_0 }  \right)^{1/2}  }$
is a universal function which describes the contribution from the 
boson anomalous dimension.
The analytic continuation gives the retarded correlation function 
in \eq{eq:DR}.

\subsection{Free energy}\label{sec:SH}

Here we compute the leading contribution to the free energy
which is generated from the quadratic action of the dressed boson,
\begin{align}
\label{eq:LogarithmBoson}
 f_B(T) = \int 
\frac{d \vec k}{ (2 \pi)^2 }
f_B(\vec k,T),
\end{align}
where $f_B(\vec k,T)$ is the contribution from the mode with momentum $\vec k$,
\begin{align}
f_B(\vec k,T) = 
\frac{3}{2}
\left(T\sum_{ \omega _m}-\int
\frac{ \dd \omega _m }{ 2 \pi }
\right)
\log\Bigl[ | \omega _m|+
\varepsilon(\vec{k}) 
\Bigr]
\end{align}
with $\varepsilon(\vec{k}) = c (|k_x|+|k_y|)$
and
$ \omega _m = 2 \pi T m$.
The thermal mass is ignored because it is higher order in $v$,
and the temperature independent ground state energy is subtracted.

Using the identity 
$ \log a =-\int^{\infty}_{0}\frac{\dd x}{x}\left(e^{-x a }-e^{-x}\right)$,
we write the free energy per mode as
\begin{align}
f_B(\vec k, T)&= 
-\frac{3}{2}
\left(T\sum_{ \omega _{m}}-\int \frac{ \dd \omega _m}{2 \pi} \right)
\int\limits^{\infty}_{0}\frac{\dd x}{x}\left(e^{-x(| \omega _m|+\varepsilon(\vec{k}))}
-e^{-x} \right).
\end{align} 
The summation over the Matsubara frequency results in
\begin{align}
f_B(\vec k,T)&= -\frac{3T}{2}\int\limits^{\infty}_{0}\frac{\dd x}{x}\left(\coth(\pi T x)-\frac{1}{\pi T x}\right)
e^{-x\varepsilon(\vec{k})}.
\end{align}
For $\varepsilon(\vec{k}) \gg T$, the free energy is suppressed only algebraically,
\bqa
f_B(\vec k,T)&= -\frac{ \pi }{2} 
\frac{ T^2 }{  \varepsilon(\vec{k}) } \left( 1  
+ O (  T/ \varepsilon(\vec{k}) )  
\right).
\eqa
This is in contrast to the non-interacting boson,
whose contribution is exponentially suppressed at large momenta.
Due to the relatively large contribution from high momentum modes,
the bosonic free energy becomes unbounded without a UV cut-off.
This leads to a violation of hyperscaling.
\begin{align}
f_B(T) \sim  - T^2 \td \Lambda,
\label{eq:fB}
\end{align}
where $\td \Lambda$ is a UV cut-off associated with irrelevant terms
as is discussed in the Appendix B.

\eq{eq:fB} is obtained without including the renormalization of the velocity 
and anomalous dimensions in \eq{eq:exponents},
which alter the scaling at intermediate energy scales.
In order to take those into account, 
we consider the scaling equation for $f_B$,
\begin{align}\label{eq:RGEq}
\left[
\left(1+\frac{2}{z}\right)-T\frac{\partial}{\partial T}+\frac{\beta_{c}}{z}\frac{\partial}{\partial c}-\frac{\td \Lambda}{z}\frac{\partial}{\partial \td \Lambda} 
\right]
f_B(T,c,\td \Lambda)=0.
\end{align}
The solution takes the form,
\begin{align}\label{eq:RGSol}
f_B(T,c,\td \Lambda)=e^{ -\int^{l}_{0}\dd l^{'} \left(1+\frac{2}{z(l^{'})}\right) }
{f}_B\left(e^{l}T, c(l), e^{ \int^{l}_{0}\frac{\dd l^{'}}{z(l^{'})}  }  \td \Lambda \right),
\end{align}
where $c(l)$ satisfies $\frac{d c(l)}{d l} = - \frac{ \beta_c }{ z( c ) }$ 
with the initial condition $c(0)=c$.
In the large $l$ limit, $z \approx 1$ and $c(l)$ is given by \eq{eq:cl}.
By choosing $l = \log 1/T$ and using the fact that 
$f_B$ is linearly proportional to $\td \Lambda$,
we obtain
\bqa
 f_B \sim \td \Lambda T^2 F_z(T).
\eqa
This is the dominant term at low temperatures
because the contribution of free electrons away from the hot spots only goes as $T^2$.
The contributions from vertex corrections are sub-leading in $v$.
Therefore, the specific heat in the low temperature limit is given by \eq{eq:c}.

\end{document}